%% file: All.tex
\documentclass[prb,twocolumn,superscriptaddress,floatfix,aps, reprint, floats]{revtex4-1}

\usepackage[english]{babel}
\usepackage[utf8x]{inputenc}
\usepackage[T1]{fontenc}

\usepackage{amsbsy,amssymb,amsmath,bm,amstext}
\usepackage{graphicx}
\usepackage[colorinlistoftodos]{todonotes}
\usepackage[colorlinks=true, allcolors=blue]{hyperref}
\usepackage{dcolumn, import}
\usepackage{transparent, color}
\usepackage{appendix}
\usepackage{upgreek}
\usepackage{gensymb}
\usepackage[mathcal]{euscript}
\usepackage{float}

\newcolumntype{L}[1]{>{\raggedright\let\newline\\\arraybackslash\hspace{0pt}}m{#1}}
\newcolumntype{C}[1]{>{\centering\let\newline\\\arraybackslash\hspace{0pt}}m{#1}}
\newcolumntype{R}[1]{>{\raggedleft\let\newline\\\arraybackslash\hspace{0pt}}m{#1}}

\begin{document}
\title{Circuit Quantum Electrodynamics of Granular Aluminum Resonators}
\author{N.~Maleeva}
\affiliation{Physikalisches~Institut,~Karlsruhe~Institute~of~Technology,~76131~Karlsruhe,~Germany}
\author{L.~Gr\"unhaupt}
\affiliation{Physikalisches~Institut,~Karlsruhe~Institute~of~Technology,~76131~Karlsruhe,~Germany}
\author{T.~Klein}
\affiliation{Universite~Grenoble~Alpes,~Institut~NEEL,~F-3800~Grenoble,~France}
\affiliation{CNRS,~Institut~NEEL,~F-3800~Grenoble,~France}
\author{F.~Levy-Bertrand}
\affiliation{Universite~Grenoble~Alpes,~Institut~NEEL,~F-3800~Grenoble,~France}
\affiliation{CNRS,~Institut~NEEL,~F-3800~Grenoble,~France}
\author{O.~Dupré}
\affiliation{Universite~Grenoble~Alpes,~Institut~NEEL,~F-3800~Grenoble,~France}
\affiliation{CNRS,~Institut~NEEL,~F-3800~Grenoble,~France}
\author{M.~Calvo}
\affiliation{Universite~Grenoble~Alpes,~Institut~NEEL,~F-3800~Grenoble,~France}
\affiliation{CNRS,~Institut~NEEL,~F-3800~Grenoble,~France}
\author{F.~Valenti}
\affiliation{Physikalisches~Institut,~Karlsruhe~Institute~of~Technology,~76131~Karlsruhe,~Germany}
\author{P.~Winkel}
\affiliation{Physikalisches~Institut,~Karlsruhe~Institute~of~Technology,~76131~Karlsruhe,~Germany}
\author{F.~Friedrich}
\affiliation{Physikalisches~Institut,~Karlsruhe~Institute~of~Technology,~76131~Karlsruhe,~Germany}
\author{W.~Wernsdorfer}
\affiliation{Physikalisches~Institut,~Karlsruhe~Institute~of~Technology,~76131~Karlsruhe,~Germany}
\affiliation{CNRS,~Institut~NEEL,~F-3800~Grenoble,~France}
\affiliation{Institute~of~Nanotechnology,~Karlsruhe~Institute~of~Technology,~76344~Eggenstein~Leopoldshafen,~Germany}
\author{A.~V.~Ustinov}
\affiliation{Physikalisches~Institut,~Karlsruhe~Institute~of~Technology,~76131~Karlsruhe,~Germany}
\affiliation{Russian~Quantum~Center,~National~University~of~Science~and~Technology~MISIS,~119049~Moscow,~Russia}
\author{H.~Rotzinger}
\affiliation{Physikalisches~Institut,~Karlsruhe~Institute~of~Technology,~76131~Karlsruhe,~Germany}
\author{A.~Monfardini}
\affiliation{Universite~Grenoble~Alpes,~Institut~NEEL,~F-3800~Grenoble,~France}
\affiliation{CNRS,~Institut~NEEL,~F-3800~Grenoble,~France}
\author{M.~V.~Fistul}
\affiliation{Russian~Quantum~Center,~National~University~of~Science~and~Technology~MISIS,~119049~Moscow,~Russia}
\affiliation{Center~for~Theoretical~Physics~of~Complex~Systems,~Institute~for~Basic~Science,~34051~Daejeon,~Republic~of~Korea}
\author{I.~M.~Pop}
\affiliation{Physikalisches~Institut,~Karlsruhe~Institute~of~Technology,~76131~Karlsruhe,~Germany}
\affiliation{Institute~of~Nanotechnology,~Karlsruhe~Institute~of~Technology,~76344~Eggenstein~Leopoldshafen,~Germany}

\begin{abstract}
The introduction of crystalline defects or dopants can give rise to so-called “dirty superconductors“ \cite{ANDERSON1959}, characterized by reduced coherence length and quasiparticle mean free path. In particular, granular superconductors\cite{Efetov2007} such as Granular Aluminum \cite{Cohen1968,Deutscher1973}  (GrAl), consisting of remarkably uniform grains connected by Josephson contacts \cite{Parmenter1967}  have attracted interest since the sixties thanks to their rich phase diagram\cite{Dynes1981,Pracht2016} and practical advantages, like increased critical temperature\cite{Abeles1966,Deutscher1973}, critical field \cite{Deutscher1977,Chui1981}, and kinetic inductance \cite{Rotzinger2016}. Here we report the measurement and modeling of circuit quantum electrodynamics\cite{Wallraff2004} properties of GrAl microwave resonators in a wide frequency range, up to the spectral superconducting gap. Interestingly, we observe self-Kerr coefficients ranging from $10^{-2}$ Hz to $10^5$ Hz, within an order of magnitude from analytic calculations based on GrAl microstructure. This amenable nonlinearity, combined with the relatively high quality factors in the $10^5$ range, open new avenues for applications in quantum information processing\cite{Nori2017} and kinetic inductance detectors\cite{Day2003}. 
\end{abstract}

\maketitle

Increasing the level of disorder in a superconducting material usually decreases the superfluid density and can induce a superconducting to insulating transition (SIT). Superconductors with low superconducting carrier density can exhibit rich physical properties, arising from a variety of phenomena such as quantum phase transitions \cite{Emery1995} and localization \cite{Efetov2007}. Granular aluminum is a typical example, preferred by experimentalists thanks to its relatively straight-forward fabrication by aluminum evaporation in an oxygen atmosphere \cite{Cohen1968}, which can tune the film resistivity $\rho$ over five orders of magnitude. The phase diagram of GrAl thin films, with an initial increase of the critical temperature versus resistivity \cite{Deutscher1973-2}, followed by a decrease and transition to an insulating state, has been extensively studied over the last fifty years, with notable recent developments in both theory \cite{Pracht2017} and experiment \cite{Bachar2015,BacharPracht2015}.  These studies, mostly performed by direct current measurements, or broad-band THz spectroscopy, offer a solid basis to start addressing the electrodynamics of GrAl in the quantum regime, defined as the limit of single photon excitations.

In the context of emerging quantum information platforms based on aluminum\cite{Nori2017}, GrAl provides precious ingredients such as low-loss and high-impedance environments, tolerance to high magnetic fields, or a robust source of nonlinearity. The prospect of implementing ultra-high impedance environments, at the level of the impedance quantum $R_Q=h/(2e) \simeq 6.5\, $ k$\,\Omega$, for the design of qubits \cite{Astafiev2012,Manucharyan2009,Gladchenko2008} and parametric amplifiers \cite{Eom2012}, or for the engineering of quantum states of light \cite{Puri2017} is very appealing. However, the electromagnetic properties of granular superconductors in the quantum regime are currently virtually unexplored.

Here we present a theoretical model and the corresponding experimental investigation of the dispersion relation and nonlinear Kerr coefficients for GrAl resonators in the microwave regime. We will use the formalism of circuit quantum electrodynamics \cite{Wallraff2004} (cQED), and show that in a first order approximation the Hamiltonian of GrAl, taking into account the interaction between the resonant modes, can be written in the familiar quantum optics form \cite{QO2008}
\begin{equation}\label{cQED_Hamiltonian}
\begin{gathered}
 \frac{H}{\hbar}=\sum_{n=1}(\omega_n +K_{nn} a_n^\dag a_n) a_n^\dag a_n +
\sum_{\substack{n,m=1 \\ n \neq m}} \frac{K_{nm}}{2}a_n^\dag a_n a_m^\dag a_m.
\end{gathered}
\end{equation}
The frequencies $\omega_n$ form the dispersion relation, the self-Kerr (sK) coefficients $K_{nn}$ quantify the frequency shift of mode $n$ for each added photon, and the cross-Kerr (cK) coefficients $K_{nm}$, quantify the frequency shift of mode $n$ for an added photon in mode $m$. The operators $a$ and $a^\dag$ are bosonic lowering and raising operators, and $a^\dag a=N$ gives the photon number.

\begin{figure*}[tbhp]
\centering
\def\svgwidth{\textwidth}  
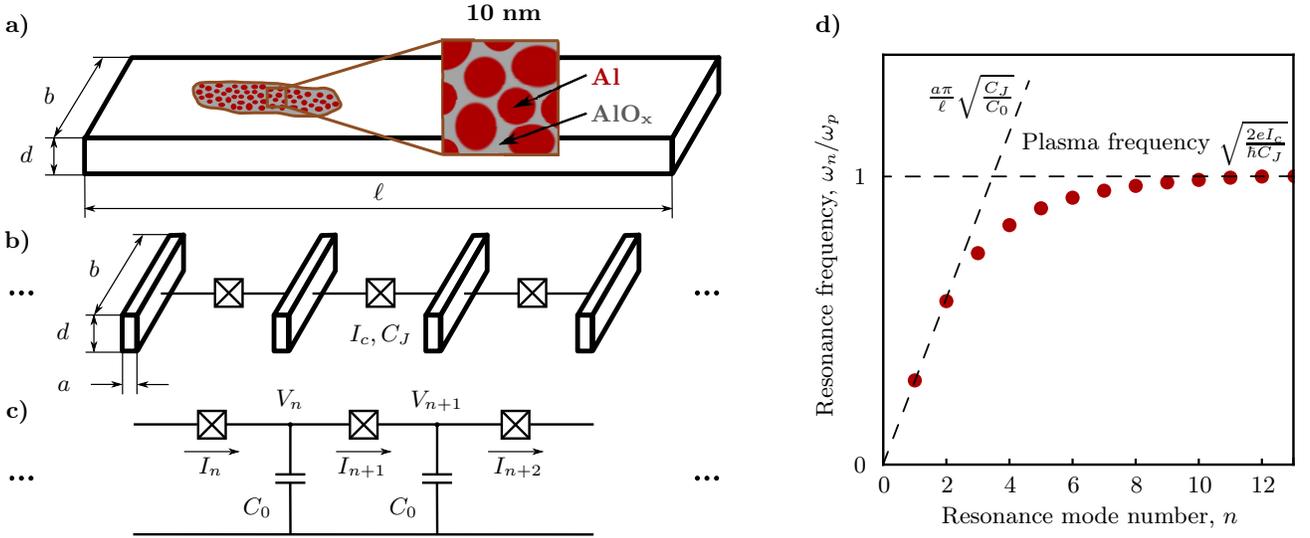
\caption{\textbf{Schematic representation of a GrAl stripline resonator with open boundary conditions.} \textbf{(a)} The length of the stripline, $\ell$, is in the range of mm, its width, $b$, is in the range of a few $\upmu$m,  and the thickness, $d$, is between 20 and 30 nm. Al grains (sketched in bordeaux color in the inset) have a diameter $a= 3\pm 1$ nm \cite{Deutscher1973}. They are separated by Aluminum oxide barriers (shown in gray), forming a 3D network of superconducting islands connected by Josephson contacts. 
\textbf{(b)} For the lowest frequency standing-current modes along the stripline, the resonator can be modeled as a 1D array of effective Josephson junctions with critical current $I_c$ and junction capacitance $C_J$, corresponding to the summed  critical currents and capacitances of the grains in a stripline section of length $a$. 
\textbf{(c)} The resulting circuit diagram consists of identical cells, each containing an effective JJ and the self capacitance $C_0$ of the superconducting island. \textbf{(d)} Typical dispersion relation of a 1D JJ array, following Eq. (\ref{omegan}).  The spectrum saturates at the effective plasma frequency $\omega_{p}= \sqrt{2e I_c / \hbar C_J}$. The slope in the linear part of the dispersion relation is defined by the ratio $\frac{a\pi}{\ell} \sqrt{C_J /C_0}$.}
\label{Schem}
\end{figure*}

\begin{figure*}[tbhp]
\centering
\normalsize
\def\svgwidth{\textwidth}  
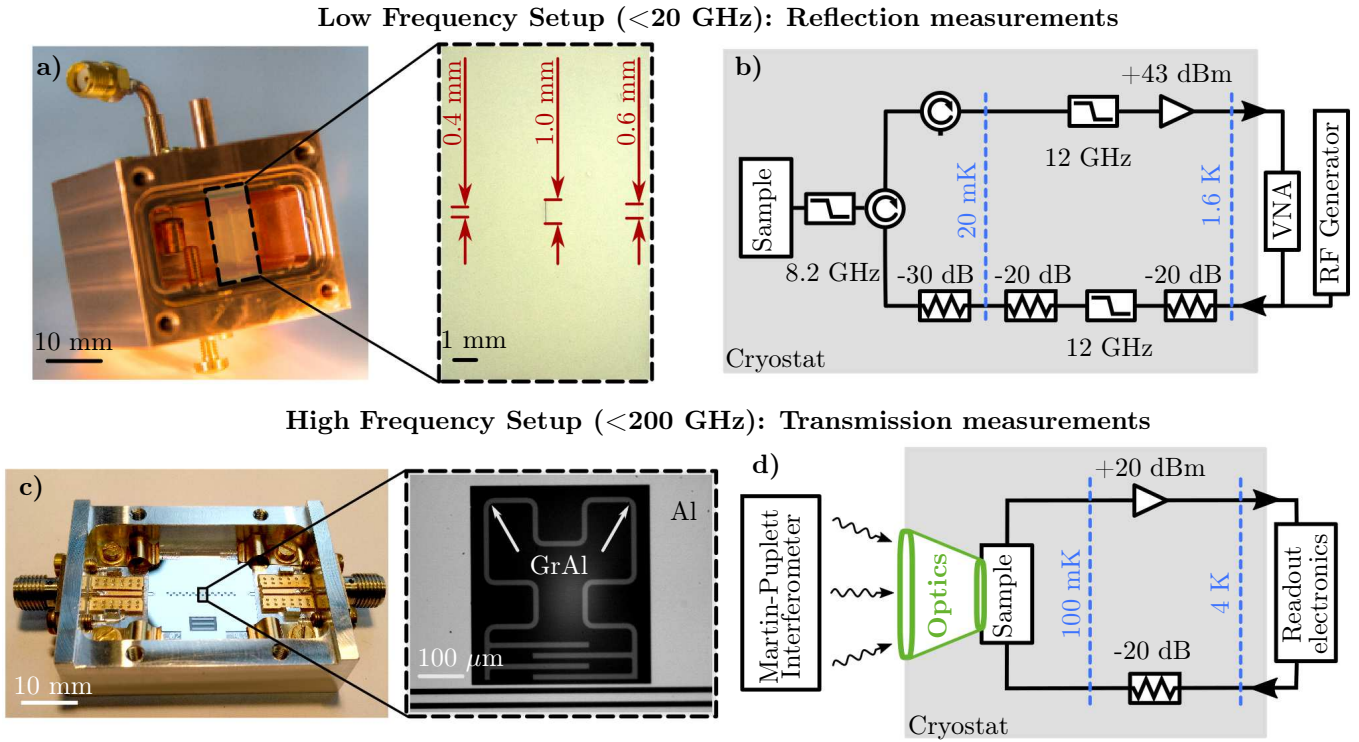
\newline
\caption{\textbf{Two complementary microwave measurement techniques for the study of GrAl resonators (GrAl$\#$1). Low frequency setup:} \textbf{(a)} Photograph of the Cu waveguide sample holder used to perform reflection measurements on stripline GrAl resonators. The inset photograph shows three of the measured resonators, with dimensions 400~$\times$~5.4~$\upmu m^2$, 600~$\times$~10~$\upmu$m$^2$, and 1000~$\times$~40~$\upmu m^2$. All resonators are 20 nm thick (see Supplementary Information). The waveguide is shielded and thermally anchored to the mixing chamber plate of a commercial dilution refrigerator. \textbf{(b)} Schematic of the cryogenic measurement setup. A reflection measurement with a vector network analyzer (VNA) characterizes the resonator response.  The total attenuation on the input lines is -70~dB, and both input, and output lines, are interrupted by commercial and custom made low pass filters providing at least -30 dB of filtering above 9~GHz. The output signal is amplified by 40~dB using a commercial high electron mobility transistor amplifier. \textbf{High frequency setup:} \textbf{(c)} Photograph of the Al sample holder and one of the resonators measured using a Martin-Puplett interferometer (MPI).  The GrAl resonators consist of a second order Hilbert shape fractal inductor and an interdigitated capacitor. Twenty two resonators are coupled to the common Al feed-line, and each resonator is surrounded by an Al ground plane. Notice the different apparent color of the GrAl film compared to Al. \textbf{(d)} Schematics of the measurement setup. The resonators are cooled down in a dilution refrigerator with optical access up to 200 GHz, facing the MPI \cite{MARTIN1970}. The optics (shown in green) consist of a lens at room temperature, and two aperture and lens pairs, at 4~K, and at 100~mK, in front of the sample \cite{Calvo2015}. The GrAl resonator response to high-frequency illumination consists in shifting its low frequency spectrum, which is continuously monitored in a transmission measurement through the common feed-line. All samples were fabricated on c-plane, double-side polished sapphire substrates, using standard e-beam and optical lithography lift-off techniques.}
\label{Setup}
\end{figure*}

\begin{figure*}[tbhp]
\centering
\def\svgwidth{\textwidth}  
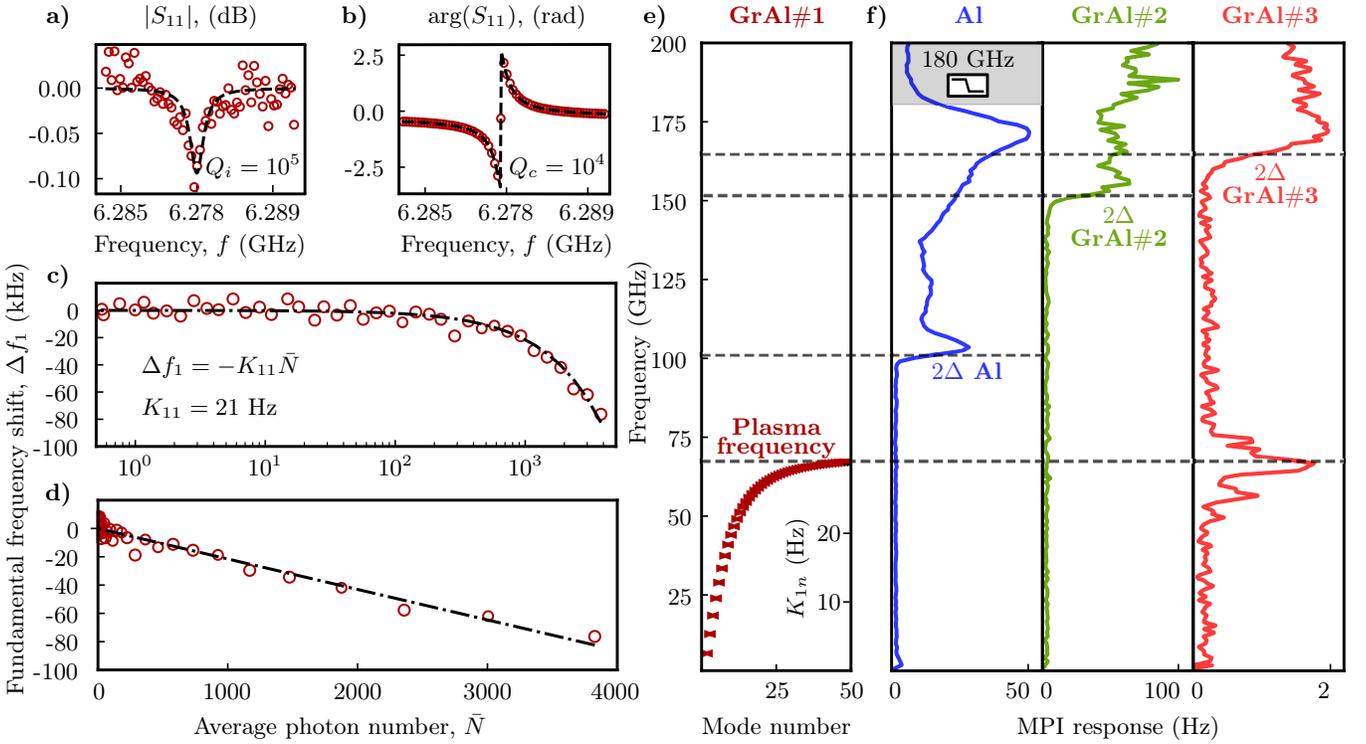
\caption{\textbf{Measurement of the dispersion relation and the nonlinearity in GrAl resonators.} Typical measured amplitude \textbf{(a)} and phase \textbf{(b)} of  the reflection coefficient $S_{11}$ for resonator GrAl\#1 (see Fig.~\ref{Setup}a). We typically observe internal quality factors of the resonators in the range of $10^5$. In \textbf{(c)} and \textbf{(d)} we plot the measured shift of the first resonant frequency versus circulating photon number $\bar{N}$ in logarithmic and linear scale, respectively. The corresponding sK coefficient extracted from the linear fit $K_{11}/2 \pi= 21$~Hz. \textbf{(e)} Calculated dispersion relation $f(n)$  for resonator GrAl\#1, starting from two tone measurements of the third mode (see text). The spectrum saturates at the effective plasma frequency 68~$\pm$~0.1~GHz. From Eq. (\ref{SelfKerr}), the cK coefficients follow the dispersion relation, and their values are reported on the right-axis. A significant cK coupling enables the observation of the high frequency spectrum, up to the effective plasma frequency: photons populating the high end of the spectrum shift the low-lying eigenfrequencies, which can be monitored via standard RF transmission measurements (see Fig.~\ref{Setup}d). \textbf{(f)} Martin-Puplett Interferometer (MPI) response of Hilbert-shaped resonators made of: 25 nm thick Al, GrAl with resistivity 80~$\upmu \Omega\,$cm (GrAl$\#$2), and GrAl with resistivity 3000~$\upmu \Omega\,$cm (GrAl$\#$3). The illumination frequencies generated by the MPI range from a few GHz up to 200 GHz, with a resolution of $1$~GHz. The different superconducting gaps of the films are evidenced by a strong MPI response due to quasiparticle excitation at 100~GHz for Al, at 150~GHz for GrAl$\#$2, and at 165~GHz for GrAl$\#$3. For the sample with the highest resistivity, and the lowest critical current density, GrAl$\#$3, we observe a peak around 65~GHz, in the  vicinity of the $\omega_p$ predicted from low frequency measurements on sample GrAl\#1, with a similarly high resistivity (4000~$\upmu \Omega\,$cm, see text for details). This MPI response can be seen as the summed dispersive frequency shift due to cK interactions $K_{1n}$ (plotted in panel e) between the fundamental mode and all higher populated modes.}
\label{All_results}
\end{figure*}

\begin{figure*}[tbhp]
\centering
\normalsize
\def\svgwidth{\textwidth}  
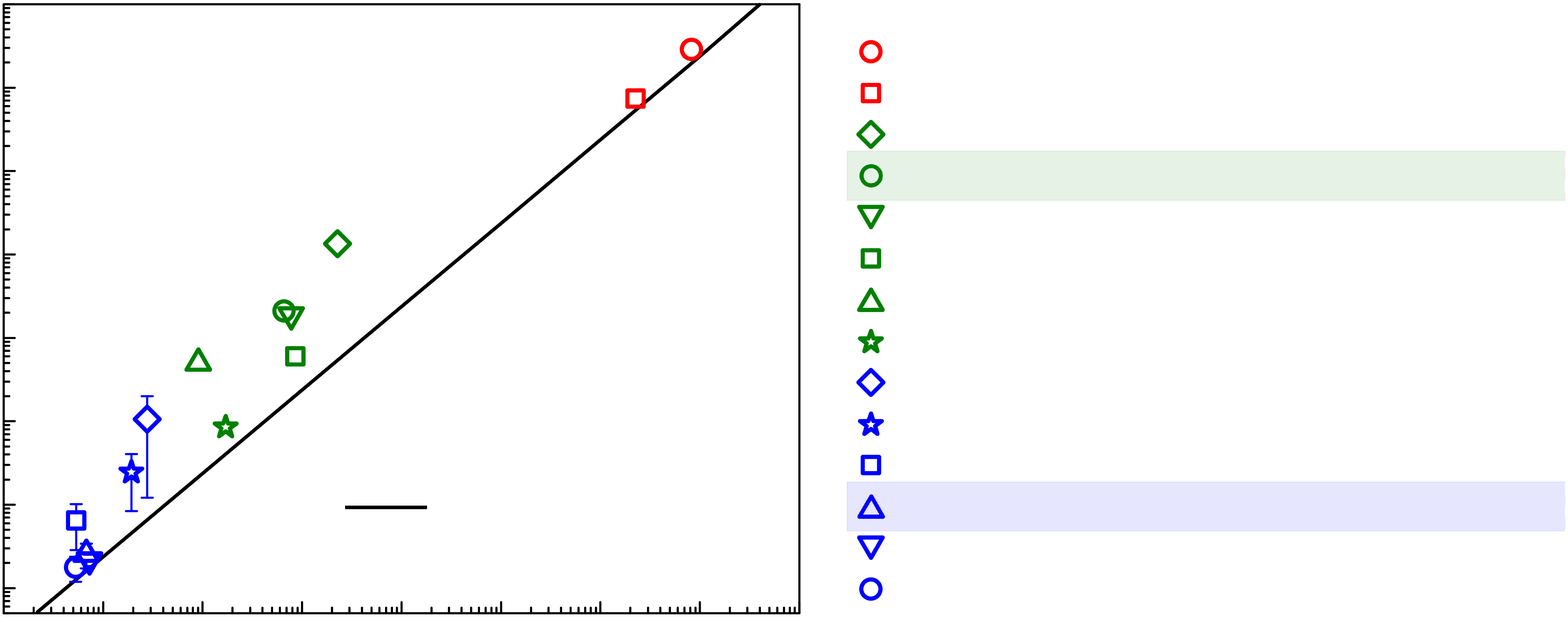
\newline
\caption{\textbf{Measured GrAl self-Kerr nonlinearity.} The measured sK coefficients of fourteen GrAl samples are plotted versus $f_1^2/j_c V_{GrAl}$, where $f_1=\omega_1/2 \pi$ is the frequency of the first mode, $j_c$ is the critical current density, and $V_{GrAl}$ is the sample volume, with values listed for each sample in the legend. Hilbert-shaped resonator samples are represented in blue, stripline samples in green, and Al-shunted stripline resonators in red. 
The error bars for the blue points show the standard deviation of the measured $K_{11}$ for nominally identical resonators. 
In the legend, the samples  are listed in decreasing $K_{11}$ order, within each group. The blue up-oriented triangle corresponds to sample GrAl$\#$2 (highlighted in blue), and the green circle corresponds to GrAl$\#$1 (highlighted in green). The black line shows the calculated sK  from Eq. (\ref{SelfKerr}), for a grain size $a=4$ nm, which includes the 1 nm thickness of the aluminum oxide barrier. We estimate the main error source to be the photon number calibration, which can only be estimated within a factor of 10. }
\label{K11}
\end{figure*}

The microstructure of GrAl consists of pure aluminum grains, with the average diameter $a$, separated by thin aluminum oxide barriers, as schematically illustrated in Fig.~\ref{Schem}a. For films fabricated at room temperature with $\rho > 10$~$\upmu \Omega \,$cm, the grain size is homogeneous and independent of resistivity, $a=3\pm 1$~nm \cite{Deutscher1973}. We use GrAl films with a resistivity between~$40$~$\upmu \Omega \,$cm~and~$4000$~$\upmu \Omega \,$cm, below the SIT at $\rho \simeq 10^4$~$\upmu \Omega \,$cm \cite{Pracht2016}, and for which the kinetic inductance dominates over the geometric inductance \cite{Rotzinger2016}. We model this medium as a network of effective Josephson junctions (JJ), which provides a handle to calculate its dispersion relation \cite{Hutter2011} and the Kerr coefficients \cite{Weissl2015,Bourassa2012}. 

For elongated structures, such as stripline resonators (Fig.~\ref{Schem}a), the calculation of the low-frequency dispersion relation and nonlinearity can be performed in the limit of one dimensional (1D) current distributions, along the stripline (see Supplementary Information), resulting in an effective JJ chain model (see Fig.~\ref{Schem}b). 
The current is homogeneously distributed through the sample cross-section, due to the fact that the thickness $d\simeq 20$~nm is much smaller than the magnetic field penetration depth, $\lambda_L>~0.4$~$\upmu$m, depending on the film resistivity $\rho$, and the width $b$ is smaller than the screening distance, $\lambda_\perp =\lambda_L^2/d >~8\ \upmu$m\cite{Cohen1968}.
The equivalent electrical  schematics is shown in Fig.~\ref{Schem}c, where each superconducting section of length $a$, with self capacitance $C_0$, is connected by effective JJs with critical current $I_c$ and capacitance $C_J$. 

The classical equation of motion for the phase difference $\varphi_{n}$  across the $n^{th}$ JJ is

\begin{equation}\label{dyn_equation}
\begin{gathered}
2I_c \sin \left( \varphi_{n+1} \right)-
I_c \sin\left( \varphi_{n+2} \right)-
I_c \sin\left( \varphi_{n} \right)+\\
+\frac{\hbar C_J}{2e}\frac{d^2 }{dt^2}\left(2\varphi_{n+1} - \varphi_{n+2} -\varphi_{n} \right)+\\
+ \delta_{m, n} I_{ext}\cos(\omega t)  =
 \frac{\hbar C_0 }{2e}\frac{d^2\varphi_n}{dt^2}.
\end{gathered}
\end{equation}
The resonator drive is introduced as an external current  applied to the $m^{th}$ cell, $\delta_{m, n} I_{ext}\cos(\omega t)$, where  $\delta_{m, n}$ is the Kronecker delta. 
In the regime of weak excitations, we can use the first order Taylor expansion for the Josephson currents (see Supplementary Information) and obtain the dispersion relation
\begin{equation}\label{omegan}
\omega_n=\frac{n a \pi}{l}\sqrt{\frac{2 e  I_c }{  \hbar \left( C_0 + \frac {n^2 \pi^2 a^2} {l^2}C_J \right)}},
\end{equation}
sketched in Fig.~\ref{Schem}d, which is approximately linear for the lowest modes, and it saturates at the effective plasma frequency $\omega_p=\omega_{n=\ell / a}~=~\sqrt{2e I_c/ \hbar C_J}$, as measured on mesoscopic JJ arrays \cite{Masluk2012}. 
As we will show in the following, the fundamental frequency $f_1=\omega_1/2 \pi$, designed in the low GHz range, can provide a convenient link through the cK effect to the higher modes of the dispersion relation, spanning up to $\sim 100$~GHz. 

To derive the Kerr coefficients of the fundamental mode in Eq.~(\ref{cQED_Hamiltonian}), we solve the equation of motion expanding the nonlinear terms up to third order. By relating the phase response amplitude to the circulating photon number $\bar{N}$ (see Supplementary Information), we obtain the sK and cK coefficients for the fundamental mode
\begin{equation}\label{SelfKerr}
K_{1n}=\mathcal{C} \pi e a \frac{ \omega_1 \omega_n}{ j_c V_{GrAl}},\ \mbox{with}\  n\geq 1.
\end{equation}
Here, $e$ is the electron charge, $a$ is the grain size, $j_c= I_c/ b d$ is the critical current density, $\omega_n$ are the eigenfrequencies given by Eq.~(\ref{omegan}), and $V_{GrAl}=b d \ell$ is the volume of GrAl threaded by current. $\mathcal{C}$ is a numerical constant of order one, which, for a sinusoidal current distribution is $\mathcal{C}=3/16$ for~$n=1$, and $\mathcal{C}=1/4$ for~$n > 1$. 
Using the expression for the single-photon current as a function of frequency and total inductance, $I_{\bar{N}=1}^2=2 f h/L$, and $L\propto 1/j_{c}$, Eq.(\ref{SelfKerr}) can be rewritten in a qualitatively similar form to the $K_{11}$ coefficient estimated from Mattis–Bardeen theory for dirty superconductors\cite{Eom2012, Rotzinger2016}, $K_{11} \propto (I_{\bar{N}=1}/I_{*})^2$. The depairing current $I_{*}$ is of the same order of magnitude as the critical current of the strip $I_c$. In contrast, Eq.(\ref{SelfKerr}) offers a quantitative model for the nonlinearity of GrAl, starting from the film properties. As we will show in Fig.~\ref{K11}, this analytic result agrees within an order of magnitude with the $K_{11}$ coefficients measured on fourteen GrAl samples, spanning from $K_{11}=2\times 10^{-2}$~Hz to $K_{11}=3\times 10^{4}$~Hz. 

Furthermore, the cK coefficients, $K_{1n}$, follow the functional dependence of the dispersion relation, $\omega_n$ given by Eq.~(\ref{omegan}), and reach a maximum at the effective plasma frequency $\omega_{p}$ (see Fig.~\ref{Schem}d and Fig.~\ref{All_results}e). Due to the high cK interaction and high mode density around $\omega_{p}$, we expect a strong response of the fundamental mode for drive frequencies in the vicinity of $\omega_{p}/ 2 \pi$. Indeed, as illustrated in  Fig.~\ref{All_results}e and f, the expected response was observed for highly resistive samples (GrAl$\#$3), with $\rho=3000~\upmu \Omega\,$cm, for which $\omega_{p}$ is low enough to be in the measurable range of the high frequency setup (Fig. \ref{Setup}d).

To measure the dispersion  relation, microwave  losses, and the nonlinearity of GrAl structures, we use three types of resonators of various shape and size (see Methods), optimized for two complementary measurement setups (see Fig.~\ref{Setup}), covering a broad frequency range up to 200 GHz.

In Fig.~\ref{All_results}a and b, we plot a typical amplitude and phase response measured for stripline resonators in the single photon regime, $\bar{N} \approx 1$, which is relevant for quantum information applications. We extract an internal quality factor $Q_{i}=10^5$, comparable to values obtained for JJ array superinductances\cite{Masluk2012}. We obtain similar results for $Q_{i}$ measurements on Hilbert-shaped (Fig.~\ref{Setup}c) and aluminum shunted stripline resonators, for tens of resonators, with GrAl resistivities up to $4000$~$\upmu \Omega \,$cm, corresponding to $\sim$k$\Omega$ characteristic impedance. As discussed in Ref.\cite{Lukas}, we estimate that $Q_{i}$ is dominated by non-equilibrium quasiparticle dissipation, which could be suppressed by phonon and quasiparticle traps.

Using a two tone spectroscopy, similar to a superconducting qubit readout procedure\cite{Wallraff2004}, we measure higher modes of the dispersion relation for stripline resonators. Due to the symmetry of the electric field, the next mode, above the fundamental, coupled to the waveguide is the third. For sample GrAl$\#$1, we measure $f_{1}=6.287$~GHz and $f_{3}=18.255$~GHz. Notice that the dispersion relation already shows a measurable deviation from linear behavior, $3\times f_{1} - f_{3} = 606\pm 1$~MHz, which, using Eq.~\ref{omegan}, allows us to estimate a effective plasma frequency $\omega_{p}= 68 \pm 0.1$~GHz (see Supplementary Information), as shown in Fig.~\ref{All_results}e. 

Indeed, using a Martin-Puplett Interferometer (MPI) as a broad-band illumination source up to 200~GHz, and a Hilbert-shaped set of resonators (GrAl$\#$3) with similar sheet resistivity as GrAl$\#$1 mounted in an optical access cryostat, we observe a strong shift of the fundamental mode for illumination frequencies in the range $60-80$~GHz (red curve in Fig.~\ref{All_results}f). We interpret this response to be the cK shift due to the population of the high mode-density region of the effective plasma frequency (see Fig.~\ref{All_results}f).

As expected, for resonators with fifty times higher critical current densities $j_c$, the effective plasma frequency can no longer be measured (green line in Fig.~\ref{All_results}f), as it is above the spectroscopic gap frequency. To confirm the correct calibration of the MPI setup, we measured the response of a standard, 25~nm aluminum film, using an additional 180~GHz low-pass filter. The MPI measurements (blue line in Fig.~\ref{All_results}e) indicate the expected Al spectral gap value of 100~GHz, above which the illumination can break Cooper pairs, inducing a shift of the fundamental mode and a $Q_i$ decrease \cite{Day2003}. Finally, notice that the spectroscopic gap of samples GrAl$\#$2 and GrAl$\#$3 increases with resistivity, as expected\cite{Pracht2016}. 

To measure the sK coefficient, $K_{11}$, we monitor the fundamental frequency as a function of photon population $\bar{N}$ using the low-frequency setup (Fig.\ref{Setup}b). Typical measurement results are shown in Fig.~\ref{All_results}c and d, in linear and logarithmic scale, respectively. In Fig.~\ref{K11}, we report the measured $K_{11}$ for fourteen types of GrAl resonators, grouped in three different geometries: KID (in blue), striplines (in green) and Al shunted striplines (in red); details on resonators geometry are given in Supplementary Information. For a direct comparison with Eq.~\ref{SelfKerr}, represented by the black line, we plot the measured sK coefficients vs. $f_1^2/j_c V_{GrAl}$, using a measured $j_c=1.1$~mA$/\upmu$m$^2$ for $\rho=1600$~$\upmu \Omega \,$cm (see Supplementary Information) and scaling it according to $j_c \propto 1/\rho$  for all resistivities \cite{Pracht2016}.  We would like to emphasize that there are no fitting parameters. We estimate the main source of the offset between the experimental points and the analytical line, to be the photon number calibration, which can only be estimated within a factor of $\sim 10$. 
Remarkably, the sK coefficient of GrAl can be tuned over six orders of magnitude by varying the room temperature resistivity $\rho \propto 1/j_{c}$ and the resonator volume $V_{GrAl}$, without compromising the internal quality factor. 

In conclusion, granular aluminum is a superconductor with high characteristic impedance, low microwave losses, and amenable nonlinearity, which recommend it as a material of choice for quantum information processing. Using a high frequency setup, including a Martin-Pupplet interferometer, we observe the effective plasma frequency of highly inductive GrAl devices in the range of 70 GHz, which is in agreement with estimates based on a 1D JJ array model and the measured low-frequency spectrum. The measured sK coefficients agree within an order of magnitude with our analytic model, and they are in the range of applications for parametrically pumped devices, such as quantum amplifiers \cite{Eom2012}. Highly inductive GrAl films could implement low-loss superinductors for quantum circuits \cite{Masluk2012} or ultra-sensitive kinetic inductance detectors \cite{CARDANI2017}. 

We are grateful to O. Buisson, G. Weiss, and A. Shnirman for fruitful discussions, and to L. Radtke and A. Lukashenko for technical support. Facilities use was supported by the KIT Nanostructure Service
Laboratory (NSL). Funding was provided by the Alexander von Humboldt foundation in the framework of a Sofja Kovalevskaja award endowed by the German Federal Ministry of Education and Research. This work was partially supported by the Ministry of Education and Science of the Russian Federation in the framework of the Program to Increase Competitiveness of the NUST MISIS, contracts no. K2-2016-063 and K2-2017-081.

\section*{Methods}
The dispersion relation for GrAl resonators spans up to $\sim 100$ GHz. To cover this wide frequency range we employ two complementary measurement setups and we use the first mode as a link between them, via the cK effect. 
The low frequency part of the spectrum ($n=1-3$), up to $20$~GHz, is measured using microwave transmission and reflection measurements in a standard cQED set-up\cite{Wallraff2004} (Fig.~\ref{Setup}b). The GrAl stripline resonators (Fig.~\ref{Setup}a) are mounted in a 3D waveguide (WG) sample holder, housed inside a hermetic copper shield coated with infrared-absorbing material. In this low-noise setup, all microwave lines are filtered above 8 GHz using commercial low-pass filters, circulators and infrared absorbers identical to the set-up in Ref. \cite{Lukas}, in order to reduce stray radiation. Even though the Hilbert-shaped GrAl resonators and their aluminum sample holder (Fig.~\ref{Setup}c) are designed to operate as kinetic inductance detectors (KIDs), which is required for the measurement of their high frequency spectrum by means of direct optical spectroscopy (Fig.~\ref{Setup}d), they were also measured by standard microwave transmission in the low-noise, shielded setup of Fig.~\ref{Setup}b. The high level of filtering and superior shielding, offered by the measurement setup optimized for low frequencies, is required for the protection of the fundamental mode against stray excitations, which is essential for the measurement of its coherence and  nonlinear properties (sK and cK).

For the measurement of the effective plasma frequency we use the wide frequency band setup of  Fig.~\ref{Setup}d, consisting of an optical access cryostat coupled to a Martin-Puplett Interferometer (see Supplementary Information). The fundamental mode is continuously measured via microwave transmission measurements, while its frequency is shifted by cK interactions with optically populated higher modes of the dispersion relation.

\bibliography{GrAl_papers.bib}

\clearpage

\onecolumngrid

\begin{center}
\huge{\textbf{Supplementary Material}}
\end{center}

\vspace*{2 cm}
\hrulefill

\section*{{A. Current distribution}}
\begin{figure*}[h!]
\centering
\def\svgwidth{0.9\textwidth}  
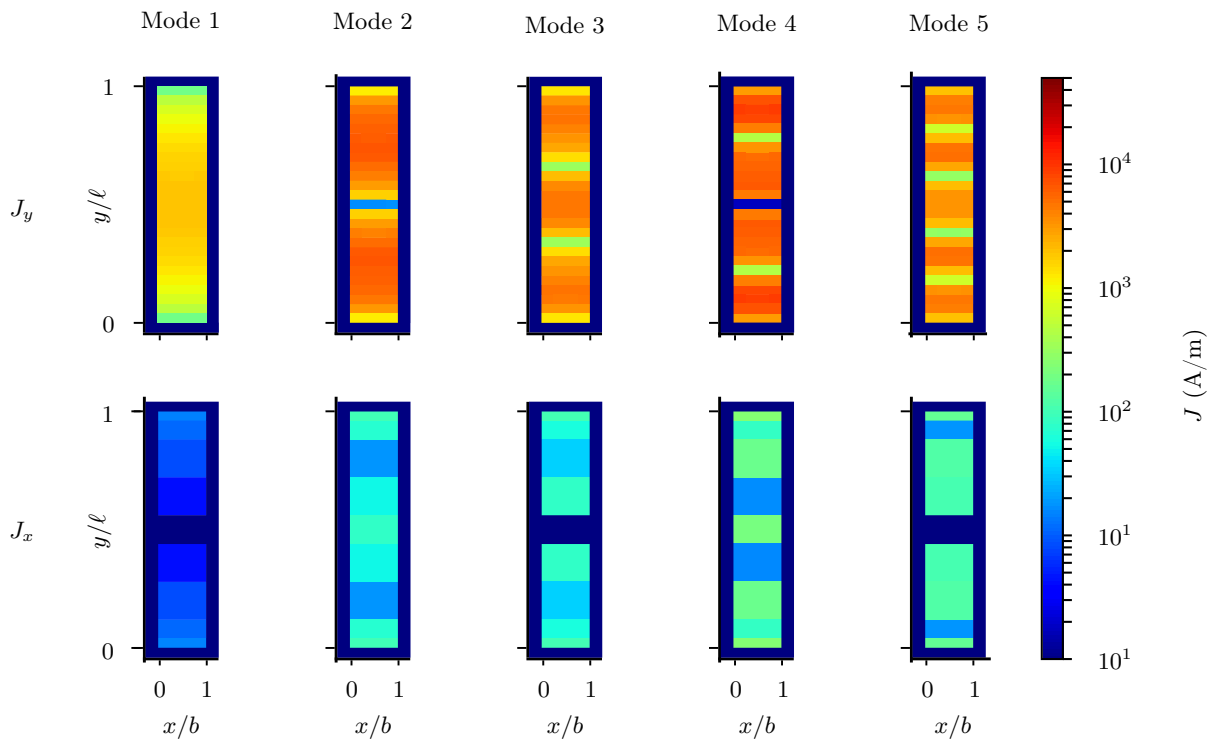
\caption{\textbf{Current distribution.} Finite elements method simulation of surface current density in a high kinetic inductance $\lambda /2$ stripline resonator with dimensions $40\times 1000$~$\upmu$m$^2$. The top row shows the current density $J_y$ along the y-dimension ($\ell =1000\ \upmu$m) for the first five modes. The corresponding current density $J_x$ along the x-dimension ($b = 40\  \upmu$m) is shown in the bottom row. $J_y$ is at least two orders of magnitude larger than $J_x$ for all modes. Therefore, we conclude that up to the fifth mode the resonator follows the well known $\lambda / 2$ current distribution, and that the measured nonlinear dispersion relation of the third mode, reported in the main text, is not caused by a drum mode like behavior.}\label{Current distr}
\end{figure*}

We perform all calculations in the limit of a one dimensional current distribution along the resonator. 
Here we present the results of a finite elements simulation of the current distribution for the five lowest stripline resonator modes. Figure~\ref{Current distr} shows the current distribution along (top row) and across (bottom row) the stripline resonator. The current along the resonator is at least two orders of magnitude higher than the current in the perpendicular direction. This validates the 1D current distribution assumption and allows us to exclude drum-like modes as a source of deviation from the linear dispersion relation.

\clearpage

\section*{{B. Details on analytical model}}\label{Analytical details}
In order to derive the equation of motion for a 1D JJ array we write the Kirchhoff laws and Josephson equations for two neighboring effective junctions as
\begin{equation}\label{1K-f}
\begin{gathered}
I_n=I_{n+1}+C_0\frac{dV_n}{dt},\\
I_{n+1}=I_{n+2} + C_0\frac{dV_{n+1}}{dt},\\
V_{n+1}-V_{n}=\frac{\hbar}{2e}\frac{d(\chi_{n+1}-\chi_{n})}{dt},
\end{gathered}
\end{equation}
where $V_n$ and $\chi_n$ are the voltage and the phase on the $n^{th}$ node respectively, and $I_n$ is the current through the $n^{th}$~JJ. Combining these equations  and introducing an excitation as an external current $I_{ext}\cos(\omega t)$ applied to the $m^{th}$ cell, where $x_m=l/2$, we obtain 
\begin{equation}\label{excRes}
2I_{n+1}-I_{n+2}-I_n+\delta_{m,n}I_{ext}\cos(\omega t) = \frac{\hbar C_0 }{2e}\frac{d^2\varphi_{n}}{dt^2},
\end{equation}
where $\varphi_{n}=\chi_{n+1}-\chi_{n}$ is the phase difference across the $n^{th}$ JJ. The JJ current is described as the following 
\begin{equation}
I_n=I_c \sin(\varphi_n) + \frac{\hbar C_J}{2e}\frac{d^2 (\varphi_n)}{dt^2}.
\end{equation}
Substituting this expression to Eq.~(\ref{excRes}) we obtain the equation of motion in the discrete limit 
\begin{equation} \label{motion eq in discreet lim}
\begin{gathered}
2I_c \sin \left( \varphi_{n+1} \right)-
I_c \sin\left( \varphi_{n+2} \right)-
I_c \sin\left( \varphi_{n} \right)+\\
+\frac{\hbar C_J}{2e}\frac{d^2 }{dt^2}\left(2\varphi_{n+1} - \varphi_{n+2} -\varphi_{n} \right)+\\
+ \delta_{m, n} I_{ext}\cos(\omega t)  =
 \frac{\hbar C_0 }{2e}\frac{d^2\varphi_n}{dt^2}.
\end{gathered}
\end{equation}
In order to obtain the dispersion relation of the resonator we rewrite Eq.~(\ref{motion eq in discreet lim}) in the continuous limit
\begin{equation}\label{Continuum}
\begin{gathered}
I_c a^2 \frac {d^2}{dx^2}\sin \varphi (x,t) 
+\frac{\hbar C_J}{2e}  a^2 \frac{d^2 }{dt^2} \frac {d^2}{dx^2}\varphi (x,t)+\\
+a\delta\left( x-\frac{\ell}{2}\right) I_{ext}\cos(\omega t)  = \frac{\hbar C_0 }{2e}\frac{d^2}{dt^2} \varphi (x,t).
\end{gathered}
\end{equation}
We consider here the first resonance mode with sinusoidal current distribution 
$
I(x,t)= I(t) \sin \left( \frac{\pi x}{\ell} \right);
$
 the corresponding phase difference is 
$
\varphi (x,t)=\varphi (t) \sin  \frac{\pi x}{\ell}.
$
By substituting the phase difference ansatz in Eq.~ (\ref{Continuum}), multiplying the equation by $\sin \frac{\pi x}{\ell}$ and integrating it along the resonator, we obtain the equation of motion of the resonator
\begin{equation}\label{Ph(t)}
\begin{gathered}
\frac{ \hbar }{2e}  \left( C_0 + \frac {\pi^2 a^2} {\ell^2}C_J \right) \frac{d^2 \varphi(t)}{dt^2}+
 2I_c  \frac {\pi^2 a^2} {\ell^2} J_1[\varphi(t)]=
 \frac{2a}{\ell} I_{ext}\cos(\omega t) .
\end{gathered}
\end{equation}
We approximate the Bessel function to first order $J_1[\varphi(t)] \sim \varphi(t)/2$, thus obtaining
\begin{equation}\label{lin}
\begin{gathered}
\alpha I_c  \frac {\pi^2 a^2} {\ell^2}   \varphi(t)+
 \frac{ \hbar }{2e} \left( C_0 + \frac {\pi^2 a^2} {\ell^2}C_J \right) \frac{d^2 \varphi(t)}{dt^2}
= \frac{2a}{\ell} I_{ext}\cos(\omega t)  .
\end{gathered}
\end{equation}
in the linear limit.
Notably this equation is very similar to the motion equation of a current biased JJ. Solving Eq. (\ref{lin}) we obtain the first resonance frequency of our system 
\begin{equation}\label{omega0}
\omega_1=\frac{ a \pi}{l}\sqrt{\frac{2 e  I_c }{  \hbar \left( C_0 + \frac {\pi^2 a^2} {l^2}C_J \right)}}.
\end{equation}
Performing the same calculations using the coordinate distribution of the higher resonance modes, we obtain the dispersion relation
\begin{equation}
\omega_n=\frac{n a \pi}{l}\sqrt{\frac{2 e  I_c }{  \hbar \left( C_0 + \frac {n^2 \pi^2 a^2} {l^2}C_J \right)}}.
\end{equation}

In order to derive the self-Kerr (sK) coefficient of the fundamental mode we solve the nonlinear equation of motion. The coupling of our resonator to the environment and internal losses of the resonator are introduced to Eq.~(\ref{Ph(t)}) as a damping term with a parameter $\gamma = \omega_1/Q_{total}$ ($\sim 10^4-10^5$ according to the experiment)
\begin{equation}\label{Nonlin}
\begin{gathered}
 \ddot{\varphi} (t) +
\frac{4 e \tilde{I_c}}{\hbar \tilde{C}} J_1[\varphi(t)] 
+\gamma \dot{\varphi}(t)= 
\frac{4 e }{ \hbar \tilde{C}} \frac{a }{ \ell} I_{ext} \cos (\omega t),
\end{gathered}
\end{equation}
where $\tilde{I_c}=  I_c \frac {\pi^2 a^2} {\ell^2}$ and $\tilde{C}= C_0 + \frac {\pi^2 a^2} {\ell^2}C_J$. Since we are working in resonance regime, we require the system to oscillate only with the driving frequency $\omega$ by assuming
\begin{equation}\label{ansatz} 
\begin{gathered}
\varphi = \varphi_a \cos(\omega t +\delta),
\end{gathered}
\end{equation}
where $\varphi_a$ is the amplitude of the response for each JJ and $\delta$ is a phase delay due to losses in the system. Solving Eq.~(\ref{Nonlin}) with ansatz (\ref{ansatz}) we obtain
\begin{equation}\label{phi_I}
\begin{gathered}
\varphi_a= \frac{\frac{4 e }{ \hbar \tilde{C}} \frac{a }{ \ell}  I }{\sqrt{\left(  \omega^2- 4 \omega_1^2  J_0[\varphi_a/2 ] J_1[\varphi_a/2 ]/\varphi_a\right)^2 + \gamma^2 \omega^2}}.
\end{gathered}
\end{equation}
For small $\varphi_a$ we can expand the Bessel functions in series up to the third order, thus obtaining
\begin{equation}\label{phi_I_1}
\begin{gathered}
\varphi_a= \frac{\frac{4 e }{ \hbar \tilde{C}} \frac{a }{ \ell}  I_{ext}}{\sqrt{ \left(\omega^2-  \omega_1^2\left(1 -  \frac{3\varphi_a^2}{32}\right)\right)^2 + \gamma^2 \omega^2}}.
\end{gathered}
\end{equation}
Since at resonance the response $\varphi_a$ reaches its highest value, we derive the resonance frequency of the nonlinear resonator by maximizing Eq.~ (\ref{phi_I_1}).
\begin{equation}\label{omega}
\omega = \omega_1\sqrt{1 -  \frac{3\varphi_a^2}{32}}\simeq \ \omega_1\left(1 -  \frac{3\varphi_a^2}{64}\right).
\end{equation}
One can see that in comparison to a single JJ with the resonance frequency, $\omega = \omega_1\left(1 -  \frac{\varphi_a^2}{4}\right)$, the 1D array has similar, but lower first order  nonlinearity. By relating the phase response to an average circulating photon number $\bar{N}$ (see Appendix~\ref{cirquit quantization}), we obtain the sK coefficient for the fundamental mode
\begin{equation}
K_{11}= \frac{3}{16} \pi ea \frac{\omega_1^2}{j_c V_{GrAl}}. 
\end{equation}

In the following, we consider the cross-Kerr (cK) coupling between two different modes $m$ and $k$ with eigenfrequencies $\omega_a$ and $\omega_b$. 
In order to obtain the cK coefficients $K_{mk}$ one needs to solve the equation of motion with excitation terms $a\delta\left( x-\frac{\ell}{2}\right)\left( I_{m}\cos(\omega_a t) + I_{k}\cos(\omega_b t) \right)$, representing the drive of $m^{th}$ and $k^{th}$ modes 
\begin{equation}\label{CrossK}
\begin{gathered}
I_c a^2 \frac {d^2}{dx^2}\sin \varphi (x,t) 
+\frac{\hbar C_J}{2e}  a^2 \frac{d^2 }{dt^2} \frac {d^2}{dx^2}\varphi (x,t)+\\
+a\delta\left( x-\frac{\ell}{2}\right)\left( I_{m}\cos(\omega_a t) + I_{k}\cos(\omega_b t) \right) = \frac{\hbar C_0 }{2e}\frac{d^2}{dt^2} \varphi(x,t).
\end{gathered}
\end{equation}
We are mainly interested in the cK coupling of the fundamental mode and we start with considering the coupling between the first and third modes. Similarly to the sK case, we look for a solution of the equation as a sum of the two driven modes 
\begin{equation}
\varphi(x,t)=\varphi_1(t)\sin\frac{\pi x}{\ell} + \varphi_3(t)\sin\frac{3\pi x}{\ell}.
\end{equation}
Using the Jacobi–Anger identity we expand the nonlinear term in Eq.~(\ref{CrossK}) up to third order 
\begin{equation}
\begin{gathered}
\sin\left( \varphi_1(t)\sin\frac{\pi x}{\ell} + \varphi_3(t)\sin\frac{3\pi x}{\ell} \right)=\\
2\sin\frac{\pi x}{\ell}\Big( J_1[\varphi_1(t)]J_0[\varphi_3(t)] + J_2[\varphi_1(t)]J_1[\varphi_3(t)] \Big)+
\\
+2\sin\frac{3\pi x}{\ell}\Big( J_3[\varphi_1(t)]J_0[\varphi_3(t)] + J_0[\varphi_1(t)]J_1[\varphi_3(t)] \Big).
\end{gathered}
\end{equation}
Here we limited the series only to the first and third modes, $\sin\left( \frac{\pi x}{\ell}\right)$ and $\sin\left( \frac{3\pi x}{\ell}\right)$, in which we are currently interested. We consider the case of strong pumping of the first mode and weak probing of the third mode. Therefore, Eq.~(\ref{CrossK}) can be simplified and splits into two separate equations for each mode
\begin{equation}\label{f1}
\begin{gathered}
\frac{\hbar }{2e}   \left(C_0+ \left(\frac{\pi a}{\ell}\right)^2C_J \right) \ddot{\varphi}_1
+2I_c \left(\frac{\pi a}{\ell}\right)^2  J_1[\varphi_1]
=
2\frac{a}{\ell}  I_{1}\cos(\omega_a t) ,
\end{gathered}
\end{equation}
\begin{equation}\label{f3}
\begin{gathered}
\frac{\hbar }{2e}   \left(C_0+ \left(\frac{3\pi a}{\ell}\right)^2C_J \right) \ddot{\varphi}_3
+I_c \left(\frac{3\pi a}{\ell}\right)^2  \Big(  1-\frac{\varphi_1^2}{4} \Big)\varphi_3 
=
-2\frac{a}{\ell}  I_{3}\cos(\omega t).
\end{gathered}
\end{equation}
Since the first mode drive is much stronger than the third mode drive, the equation of motion of the first mode (\ref{f1}) contains only sK nonlinearity and looks like Eq.~(\ref{Nonlin}). On the contrary, the equation of motion of the third mode (\ref{f3}) is linear in $\varphi_3$ and contains only cK nonlinearity.
In analogy to the sK coefficient derivation, we look for the solution of Eq.~(\ref{f3}) at the same frequency of the drive $\omega$
\begin{equation}\label{ansatz3}
\varphi_3=\varphi_b\cos(\omega t+\delta),
\end{equation} 
where $\varphi_a$ is the amplitude of the third mode response for each JJ  and $\delta$ is a phase delay due to losses in the system. Solving Eq.~(\ref{f3}) with ansatz (\ref{ansatz3}) we obtain
\begin{equation}\label{phi_I_3}
\begin{gathered}
\varphi_b= \frac{\frac{4 e }{\hbar\left(C_0+ \left(\frac{3\pi a}{\ell}\right)^2C_J \right)} \frac{a }{ \ell}  I_3}
{\sqrt{ \left(\omega^2-  \omega_3^2\left(1 -  \frac{\varphi_a^2}{8}\right)\right)^2 + \gamma^2 \omega^2}}.
\end{gathered}
\end{equation}
Again, we derive the resonance frequency by maximizing Eq.~(\ref{phi_I_3})
\begin{equation}
\omega = \omega_3\sqrt{1 -  \frac{\varphi_a^2}{8}}\simeq \ \omega_3\left(1 -  \frac{\varphi_a^2}{16}\right),
\end{equation}
which gives the cK coefficient between the first and third modes
\begin{equation}
K_{13}=\frac{1}{4} \pi ea \frac{\omega_1 \omega_3}{j_c V_{GrAl}}.
\end{equation}
Performing the same procedure for first and all the other modes, we obtain the cK coefficients 
\begin{equation}
K_{1n}=\frac{1}{4} \pi ea \frac{\omega_1 \omega_n}{j_c V_{GrAl}}.
\end{equation}

\clearpage

\section*{{C. Circuit quantization}}\label{cirquit quantization}

The total $Q_{total}$ and coupling quality factor $Q_c$ can be extracted from the measurement, allowing the average number of photons in the resonator to be calculated. In the case of one port waveguide for the first resonance mode it can be written as 
\begin{equation}
\bar{N}=P_{in} \frac{4Q_{total}^2}{\hbar \omega_1^2 Q_c},
\end{equation}
where $P_{in}$ is the input power at the cavity port. The average number of photons relates to the amplitude of the current circulating in the resonator as \begin{equation}\label{Ires versus N}
I_{res}^2=2 \pi \frac{a}{\ell} \hbar \omega_1  \bar{N}/ L_J , 
\end{equation} 
where $L_J=\frac{\hbar}{2e I_c}$ is the Josephson inductance of one junction \cite{Devoret2004}.
At resonance, $\varphi_a$ reaches its maximal value, which is 
\begin{equation}\label{phi}
\varphi_{a}=\frac{2\pi }{\Phi_{0}} I_{res} L_{J},
\end{equation}
where $\Phi_0=h/2e$ is the (superconducting) magnetic flux quantum. The sK nonlinearity is proportional to the second order of the phase response
\begin{equation}\label{phi_a^2}
\varphi_a^2=4\pi e a \frac{\omega_1}{j_c V_{GrAl}} \bar{N}
\end{equation}

Substituting Eq. (\ref{phi_a^2}) in Eq. (\ref{omega}) we obtain
\begin{equation}
\begin{gathered}
\omega = \omega_1 -  \frac{3 }{16} \frac{\pi e \omega_1^2}{I_c}\frac{a}{\ell}\bar{N}=\omega_1-K_{11}\bar{N}
\end{gathered}
\end{equation}

\clearpage
\section*{{D. Details on samples}}\label{App Samples}

\textbf{This section refers to figures in the main text when not specified otherwise.} 
We measure the fundamental frequency of all resonators directly with the VNA. Due to its symmetry, the second mode is decoupled from the waveguide mode. For the two longest resonators made of GrAl$\#$1 film (4000 $\upmu \Omega\,$cm), the third mode, which is outside the frequency range of our VNA, can be excited by a second tone, generated by an RF generator, and detected via its cK interaction with the first mode. Knowing the first and third resonance frequencies we use Eq.~(3) of the main text to derive the plasma frequency
\begin{equation}
\omega_p=2\omega_1 \omega_3\sqrt{\frac{2}{9\omega_1^2-\omega_3^2}},
\end{equation}
which for $f_{1}=6.287\pm 0.001$~GHz and $f_{3}=18.255\pm 0.001$~GHz gives $\omega_{p}= 68 \pm 0.1$~GHz.

Figure~3a and b depict the measured $S_{11}$ response in the single photon regime for the stripline resonator of 0.6 mm length and 0.04 mm width made of GrAl$\#$1 film. A circle fit routine \cite{Probst2015} is used to extract the internal and coupling quality factors $Q_i=10^5$ and $Q_c=10^4$ respectively \cite{Lukas}, and the resonance frequency $f_1\approx 6.3$~GHz. 

\begin{center}
\begin{table*}[tbhp]
\caption{Details on the samples from Fig.~4}\label{Samples_Table}
\begin{tabular}{|C{0.8in}|C{0.8in}|C{0.8in}|C{0.8in}|C{0.8in}|C{0.8in}|C{0.8in}|}
$\rho$, $\upmu \Omega\,$cm & $\ell$, $\upmu$m & $b$, $\upmu$m & $V_{GrAl}$, $\upmu$m$^3$ & $f_1$, GHz & $f_3$, GHz & $K_{11}^{exp}$, Hz  \\
 \hline 
2000 & 2.7  & 0.05  & 0.003& 4.7031  &   -   &29$\times10^3$  \\ 
2000 &  2.7 &  0.2  & 0.01 &  4.9555 &    -  & 7.4$\times10^3$ \\
\hline
4000 &  400  &  5.4 & 43.2 & 6.995   & 	-	 &  135  \\ 
4000 &  600  &  10  & 120  & 6.287   & 18.255  &  21  \\
2800 &  600  &  8.9 &  107 & 7.6139  &   -   & 18.8 \\
2800 &  600  &  7.3 &  88  & 7.231   &   -   & 6 \\
4000 & 1000  &  40  & 800  & 6.024   & 17.645  &   5 \\
2800 & 1000  & 31.2 &  624 & 8.635   &   -   & 0.9 \\
\hline
1600 &  400  &  5.4 & 288  & 3.16 &   -  &  1.1  \\ 
900  &  600  &  10  & 288  & 3.51 &  -    &  0.2  \\
40   &  600  &  8.9 &  100 & 5.18 &   -   & 65$\times10^{-3}$ \\
  80 &  600  &  7.3 &  100 & 4.12 &   -   & 26$\times10^{-3}$ \\
220  & 1000  &  40  & 100  & 2.58 &   -   & 22$\times10^{-3}$  \\
160  & 1000  & 31.2 &  100 & 2.57 &   -   & 18$\times10^{-3}$ \\
\hline
\end{tabular}
\end{table*}
\end{center}

\begin{figure*}[tbhp]

\centering
\def\svgwidth{\textwidth}  
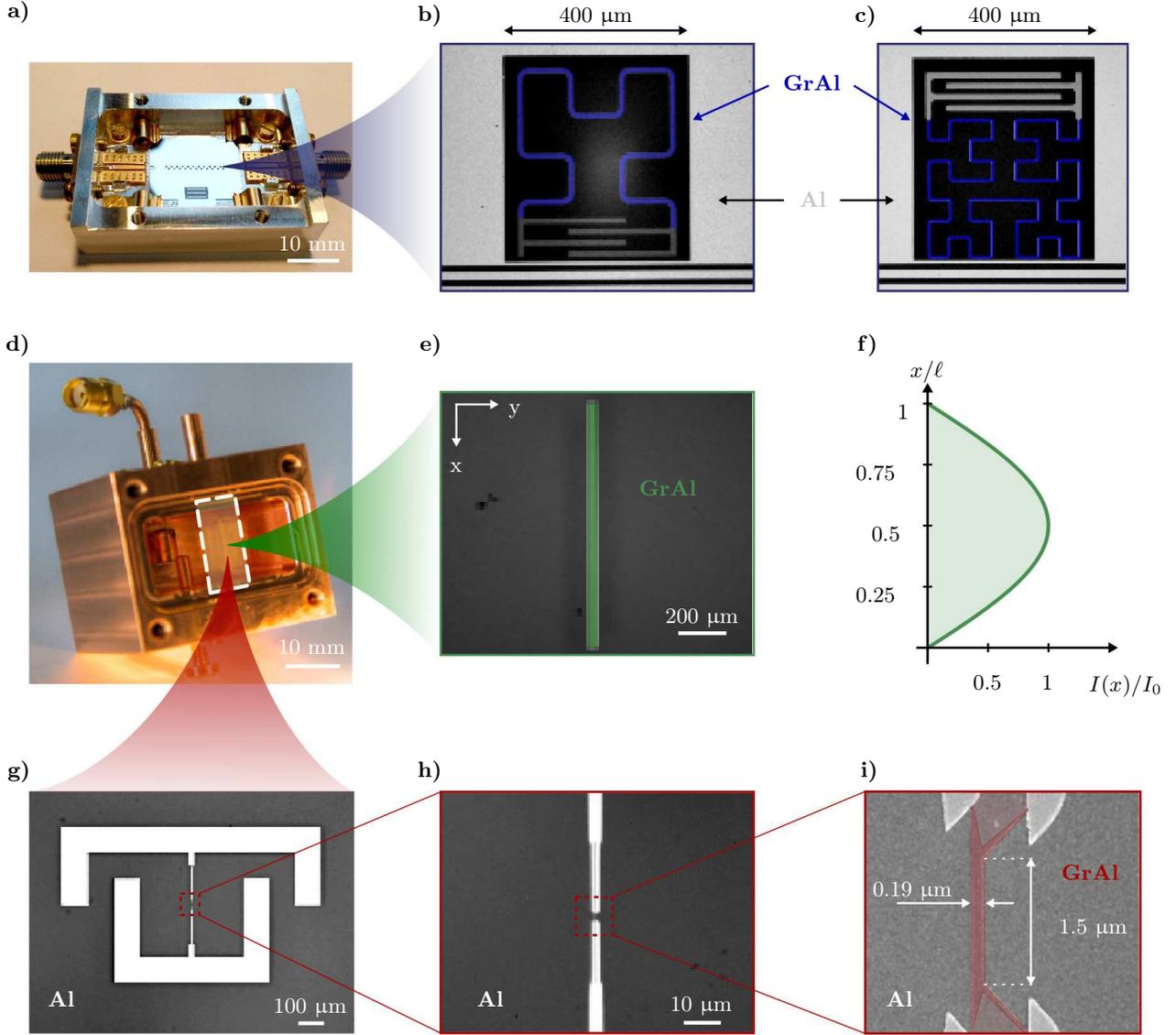
\caption{\textbf{(a)} 2D sample holder for reflection and transmission measurements of microwave resonators which are coupled to an on-chip feedline in a notch-type geometry. \textbf{(b)}, \textbf{(c)} Lumped element resonators fabricated from granular aluminum (GrAl) in a standard KID geometry with inductors in a Hilbert 2 and Hilbert 3 meander shape, respectively, and interdigitated shunt capacitors. The resonators are either coupled capacitively or inductively to the aluminum CPW transmission line. For both samples, the GrAl volume contributing to the total kinetic inductance of the fundamental mode and, thus to the sK coefficient $K_{\mathrm{11}}$, is highlighted in blue. Although the inductor in the Hilbert 3 shape is much longer, the smaller wire width results in an overall smaller total volume. \textbf{(d)} 3D sample holder for reflection measurements. \textbf{(e), (f)} Rectangular shaped distributed microstrip-stripline resonator with open boundary conditions fabricated from GrAl. Although the whole resonator volume contributes to the kinetic inductance (green shaded area), the weighting of each volume element is determined by the standing wave current distribution of the fundamental mode along the resonator. \textbf{(g)} Aluminum-shunted lumped element GrAl resonator formed by two large aluminum islands connected via a thin bridge. \textbf{(h)} The kinetic inductance of pure aluminum and the geometric inductance of the bridge are neglectable, only a small un-shunted volume of highly resistive GrAl in the center of the bridge is contributing to the total inductance of the resonator. \textbf{(i)} The zoom-in into the bridge center shows the un-shunted GrAl film, shaded in light red, and the total GrAl volume contributing to the kinetic inductance according to the expected current flow, shaded in dark red.}\label{samplespatrick}
\end{figure*}

An overview of samples holders and sample geometries is provided in Fig.~\ref{samplespatrick}  (in this appendix).

\clearpage

\section*{{E. Martin-Puplett Interferometer}}\label{App MPI}

\begin{figure*}[tbhp]
\includegraphics[width=\textwidth]{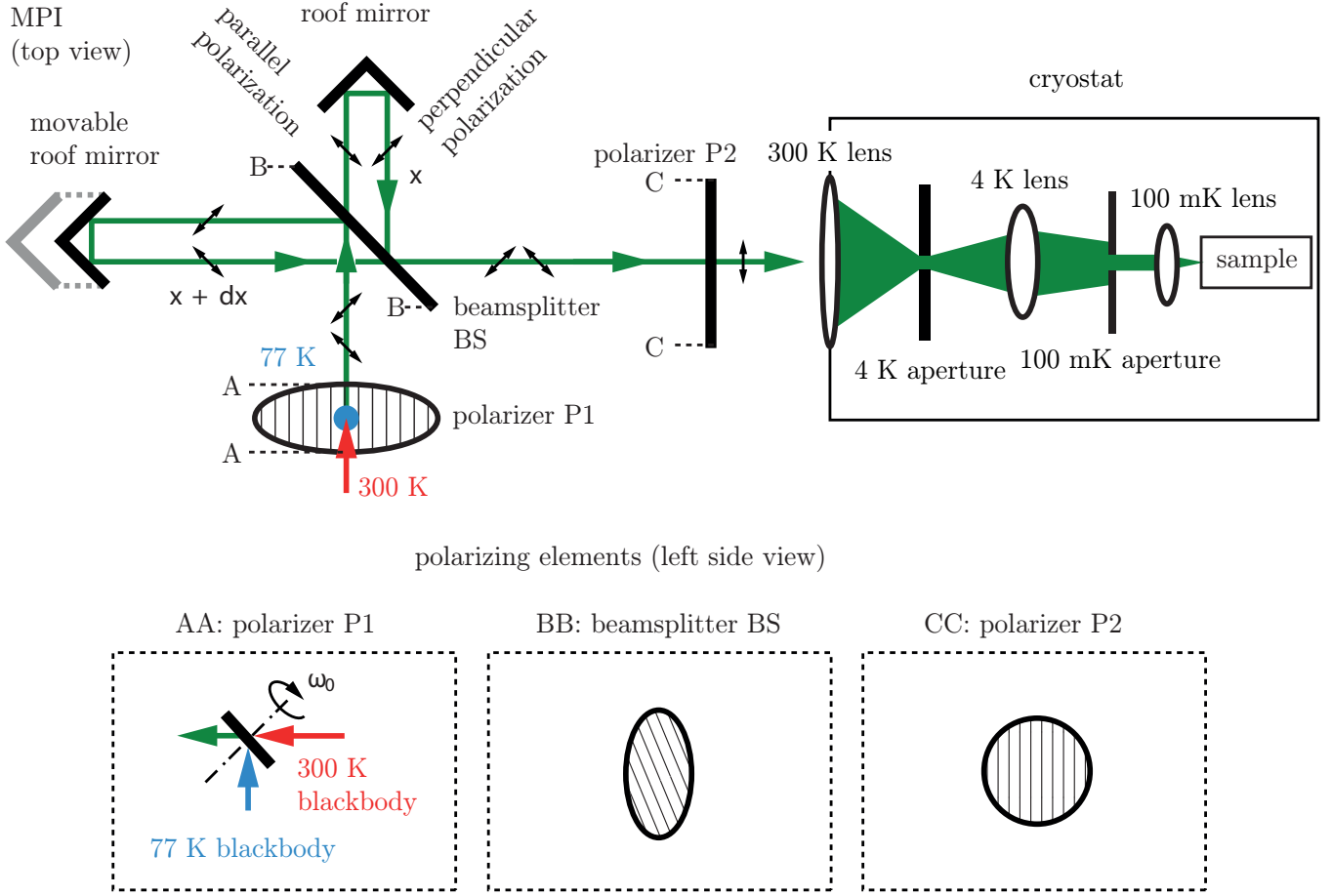}
\caption{\textbf{Optical diagram of the Martin-Puplett interferometer.} The source radiation is two combined beams from two blackbodies at 77 K (liquid Nitrogen) and 300 K. The rotating polarizer P1 combines and polarizes the beam, which is then divided by two partial beams by the beam splitter BS. The wire grids of the beam splitter BS are oriented at 45$^{\degree}$ to the normal of the drawing so that the polarization component perpendicular to the grid is transmitted and the component parallel to the grid is reflected. Two roof mirrors (one of each is movable) bring two components back to BS with a 90$^{\degree}$ rotation of the polarization. If the two roof mirrors are equally spaced from BS, the input and output beams of the BS are polarized identically. After polarizer P2 only one orthogonal polarization is transmitted to the cryostat. The cryostat optical system consists of a lens at room temperature, and two aperture and lens pairs, at 4~K, and at 100~mK, in front of the sample. } 
\label{MPI scheme}
\end{figure*}

We use the Martin-Puplett interferometer (MPI) as a broad-band illumination source, with a resolution up to $1$ GHz. 
A schematic drawing of the MPI is shown in Fig.~\ref{MPI scheme}. Three wire grids are used as polarizer P1, beam splitter BS and polarizer P2. If the polarization of an incident wave is parallel to the wires, the wire grid behaves like the surface of a metal and the wave is reflected, whereas it would be a perfectly transparent element for a wave that is polarized orthogonally to the wires. The source radiation, emitted from two black bodies, at room temperature (red) and liquid nitrogen temperature (blue), is combined on the polarizer P1, which is a wire grid inclined by $45 ^{\circ}$ with respect to the plane of the drawing. The polarizer P1 is rotating with frequency $\omega_0$ about its axis, which creates the output beam containing two orthogonal polarizations that are swept over all possible orientations by a full rotation of P1.
In the following discussion we'll consider P1 at a fixed moment in time.
The beam is divided into two partial beams by the beam splitter BS, which is a fixed wire grid at a $45 ^{\circ}$ angle to the incoming beam. The component of the incident beam with polarization parallel to the wires of BS is transmitted towards the fixed roof mirror, and the orthogonal component is reflected towards the movable roof mirror. The roof mirrors reflect the incident beams and flip their polarizations by $90 ^{\circ}$. The movable roof mirror can be displaced by an amount $dx$. The two beams recombine at BS with an accumulated path difference $\Delta = 2d x$, resulting in a phase difference $2\pi \Delta / \lambda $ which gives rise to interference. After the reflection both polarizations are flipped with respect to their first encounter with BS, therefore the transmission/reflection routine is inverted. The polarizer P2 provides a reference for the $\omega_0$ modulation of the polarization produced by P1. The single polarization output beam enters the cryostat through the room temperature lens. Here, the incoming beam is collected and sent to the cold optics, consisting on two additional focusing lenses, an in-focus aperture to reduce the intensity, and an out-of-focus aperture to crop the image.
The intensity of the on-sample radiation for a single wavelength is \cite{MARTIN1970}
\begin{equation} \label{intensity}
I(\Delta) \propto I_0(\lambda) (1+\text{cos}(2\pi \Delta  / \lambda )),
\end{equation}
where $I_0$ is defined among others by the diameter of the apertures and position of the sample with respect to the optical exes of cryostat optical system.
$I(\Delta)$ is an even, periodic function of the roof mirror displacement, with the first maximum appearing at the origin, which is result of the alternatingly constructive and destructive interference.

\begin{figure*}[tbhp]
\centering
\def\svgwidth{0.97 \textwidth}  
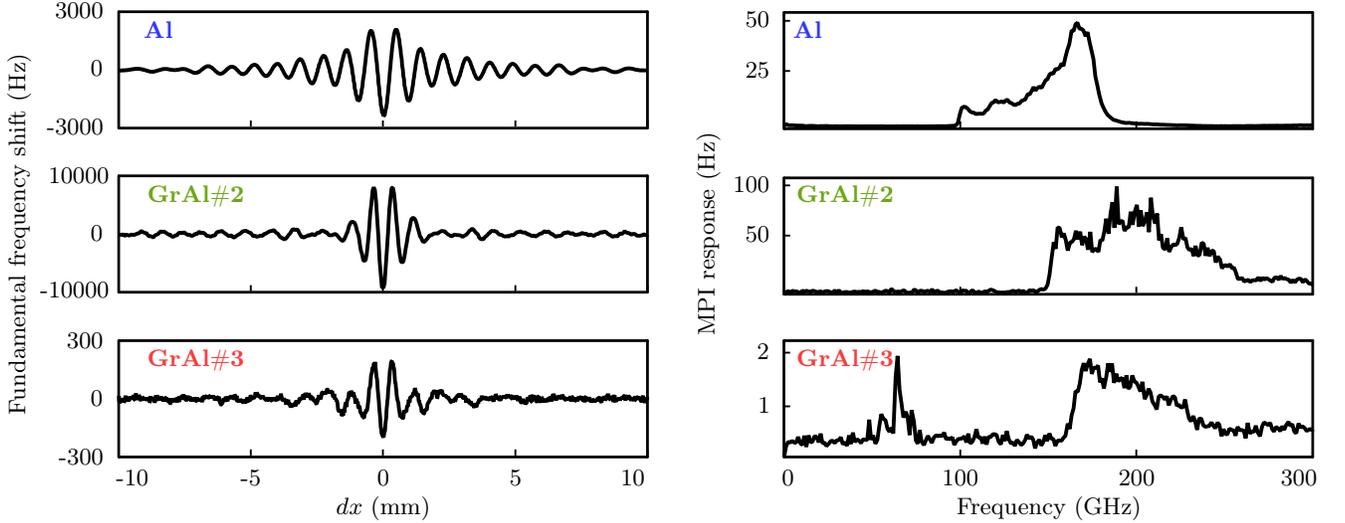
\caption{Interferograms (left) and relative spectra (right, also present in the main text) of Hilbert-shaped resonators made of Al, GrAl$\#$2, and GrAl$\#$3. The black line shows the averaging over for nominally identical resonators at each sample.} 
\label{MPI response}
\end{figure*}

 The interferogram is the modulated term of Eq.~(\ref{intensity}) integrated over all wavelengths. 
The parity of the integrand allows us to recast the cosine modulation as an exponential having the same argument, thus showing that the interferogram and the spectrum are Fourier transform (FT) pairs. The interferogram has a global maximum at the origin since different wavelengths will interfere in a fully constructive fashion only in the case of zero path difference. Furthermore, the interferogram is a stronger signal than the monochromatic intensity, since it encodes contributions from all wavelengths. Interferograms and their relative spectra are shown in Fig.~\ref{MPI response}.

The interferogram is generated by recording the on-sample irradiation at roof mirror steps $T_x \sim 10\ \upmu$m. This is equivalent to multiplying the a priori continous signal with a comb of Dirac deltas with spacing $T_x$. The FT of this product is a convolution of the spectrum with a $1/T_x$ Dirac comb in impulse space, i.e. an array composed by images of the spectrum spaced $1/T_x$ apart. The images are symmetrical and bounded by some $\pm f_\text{max}$. The highest frequency that can be attained before image overlapping and subsequent aliasing is given by step $T_x$, for $T_x\approx 50\ \upmu$m the highest frequency $f_{max}=c/2T_x \approx 3000$ GHz. For all measurements presented in Fig.~\ref{MPI response} lowpass filters are used, for Al samples the cutoff frequency is 180 GHz, for GrAl$\#$2 and GrAl$\#$3 the cutoff frequency is 300 GHz

\clearpage

\section{{Critical current}}

\begin{figure}[tbhp]
\centering
\def\svgwidth{\columnwidth}  
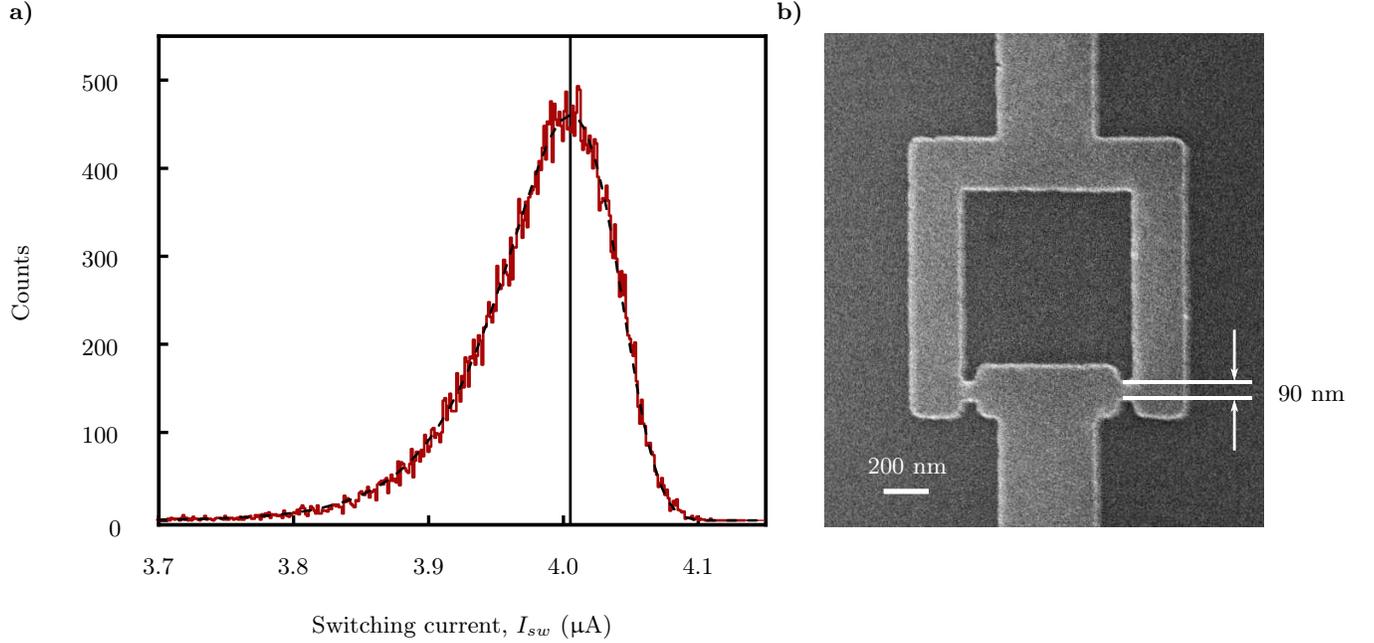
\caption{\textbf{Switching current distribution of a GrAl DC-SQUID. (a)} Switching current histogram at effective zero field for the SQUID shown in (b). The red curve shows the measured data. For each measurement, the bias current through the SQUID is increased at a constant rate until a finite voltage drop across the SQUID is detected. The applied current at this switching point defines $I_\mathrm{sw}$. From a fit to the data (black dashed line) according to the expected switching probability distribution a mean value of $4.01\,\upmu\mathrm{A}$ is obtained (for details on the fitting curve see \cite{bluhm2008deconvolution,
kurkijarvi1972intrinsic}).  \textbf{(b)} SEM image of the respective SQUID. The sample design was patterned on a $\mathrm{Si/SiO}_2$ wafer by e-beam lithography followed by the evaporation of a 20\,nm thick GrAl thin film with a sheet resistance of $1600$ $\upmu \Omega \cdot $cm. The two SQUID junctions have a combined cross section of $2\times 90\,$nm$\times 20\,$nm, resulting in a critical current density $j_\mathrm{c} \approx 1.1\,$mA$/\upmu \mathrm{m}^2$, quoted in the main text.}\label{graldcsquid}
\end{figure}

The geometry and the switching current measurements for a DC-SQUID made with granular aluminum are presented in Fig~\ref{graldcsquid}.

\end{document}

%% file: Schem5.eps_tex
\begingroup%
  \makeatletter%
  \providecommand\color[2][]{%
    \errmessage{(Inkscape) Color is used for the text in Inkscape, but the package 'color.sty' is not loaded}%
    \renewcommand\color[2][]{}%
  }%
  \providecommand\transparent[1]{%
    \errmessage{(Inkscape) Transparency is used (non-zero) for the text in Inkscape, but the package 'transparent.sty' is not loaded}%
    \renewcommand\transparent[1]{}%
  }%
  \providecommand\rotatebox[2]{#2}%
  \ifx\svgwidth\undefined%
    \setlength{\unitlength}{899.10015917bp}%
    \ifx\svgscale\undefined%
      \relax%
    \else%
      \setlength{\unitlength}{\unitlength * \real{\svgscale}}%
    \fi%
  \else%
    \setlength{\unitlength}{\svgwidth}%
  \fi%
  \global\let\svgwidth\undefined%
  \global\let\svgscale\undefined%
  \makeatother%
  \begin{picture}(1,0.39822528)%
    \put(0,0){\includegraphics[width=\unitlength]{Schem5.eps}}%
    \put(0.15077898,0.04982754){\color[rgb]{0,0,0}\makebox(0,0)[b]{\smash{$I_n$}}}%
    \put(0.26293107,0.04989418){\color[rgb]{0,0,0}\makebox(0,0)[b]{\smash{$I_{n+1}$}}}%
    \put(0.37861561,0.04989418){\color[rgb]{0,0,0}\makebox(0,0)[b]{\smash{$I_{n+2}$}}}%
    \put(0.20955102,0.09990181){\color[rgb]{0,0,0}\makebox(0,0)[b]{\smash{$V_n$}}}%
    \put(0.31837727,0.09956852){\color[rgb]{0,0,0}\makebox(0,0)[b]{\smash{$V_{n+1}$}}}%
    \put(0.04133623,0.1489373){\color[rgb]{0,0,0}\makebox(0,0)[b]{\smash{$d$}}}%
    \put(0.06447047,0.19575127){\color[rgb]{0,0,0}\makebox(0,0)[b]{\smash{$b$}}}%
    \put(0.01416188,0.28062452){\color[rgb]{0,0,0}\makebox(0,0)[b]{\smash{$d$}}}%
    \put(0.03097058,0.3239622){\color[rgb]{0,0,0}\makebox(0,0)[b]{\smash{$b$}}}%
    \put(0.04194985,0.11093655){\color[rgb]{0,0,0}\makebox(0,0)[b]{\smash{$a$}}}%
    \put(0.27530605,0.25303701){\color[rgb]{0,0,0}\makebox(0,0)[b]{\smash{$\ell$}}}%
    \put(0.18536372,0.01904856){\color[rgb]{0,0,0}\makebox(0,0)[b]{\smash{$C_0$}}}%
    \put(0.27685138,0.14770323){\color[rgb]{0,0,0}\makebox(0,0)[b]{\smash{$I_c,C_J$}}}%
    \put(0.43283877,0.33811584){\color[rgb]{0.6745098,0,0}\makebox(0,0)[lb]{\smash{$\bf{Al}$}}}%
    \put(0.43274981,0.31142248){\color[rgb]{0.35294118,0.35294118,0.35294118}\makebox(0,0)[lb]{\smash{$\bf{AlO_x}$}}}%
    \put(0.3681807,0.38619623){\color[rgb]{0,0,0}\makebox(0,0)[b]{\smash{$\bf{10\ nm}$}}}%
    \put(0.2939167,0.01904856){\color[rgb]{0,0,0}\makebox(0,0)[b]{\smash{$C_0$}}}%
    \put(0.650339,0.03559114){\color[rgb]{0,0,0}\makebox(0,0)[b]{\smash{0}}}%
    \put(0.69718923,0.03559114){\color[rgb]{0,0,0}\makebox(0,0)[b]{\smash{2}}}%
    \put(0.74403946,0.03559114){\color[rgb]{0,0,0}\makebox(0,0)[b]{\smash{4}}}%
    \put(0.79088963,0.03559114){\color[rgb]{0,0,0}\makebox(0,0)[b]{\smash{6}}}%
    \put(0.8377398,0.03559114){\color[rgb]{0,0,0}\makebox(0,0)[b]{\smash{8}}}%
    \put(0.88459003,0.03559114){\color[rgb]{0,0,0}\makebox(0,0)[b]{\smash{10}}}%
    \put(0.93144015,0.03559114){\color[rgb]{0,0,0}\makebox(0,0)[b]{\smash{12}}}%
    \put(0.802787,0.01252381){\color[rgb]{0,0,0}\makebox(0,0)[b]{\smash{Resonance mode number, $n$}}}%
    \put(0.61215146,0.20773572){\color[rgb]{0,0,0}\rotatebox{90}{\makebox(0,0)[b]{\smash{Resonance frequency, $\omega_n/ \omega_{p}$ }}}}%
    \put(0.85083498,0.2901347){\color[rgb]{0,0,0}\makebox(0,0)[b]{\smash{Plasma frequency $\sqrt{\frac{2eI_c}{\hbar C_J}}$}}}%
    \put(0.71451916,0.32461346){\color[rgb]{0,0,0}\makebox(0,0)[b]{\smash{$\frac{a \pi}{\ell} \sqrt{\frac{C_J}{C_0}}$}}}%
    \put(0.98585849,0.20773572){\color[rgb]{0,0,0}\rotatebox{90}{\makebox(0,0)[b]{\smash{ }}}}%
    \put(0.00783634,0.37734891){\color[rgb]{0,0,0}\makebox(0,0)[b]{\smash{\textbf{a)}}}}%
    \put(0.60932664,0.37734891){\color[rgb]{0,0,0}\makebox(0,0)[b]{\smash{\textbf{d)}}}}%
    \put(0.63254343,0.050451){\color[rgb]{0,0,0}\makebox(0,0)[b]{\smash{0}}}%
    \put(0.63257602,0.26460992){\color[rgb]{0,0,0}\makebox(0,0)[b]{\smash{1}}}%
    \put(0.00783634,0.21718878){\color[rgb]{0,0,0}\makebox(0,0)[b]{\smash{\textbf{b)}}}}%
    \put(0.00783634,0.09084023){\color[rgb]{0,0,0}\makebox(0,0)[b]{\smash{\textbf{c)}}}}%
  \end{picture}%
\endgroup%

%% file: All_exp2.eps_tex
\begingroup%
  \makeatletter%
  \providecommand\color[2][]{%
    \errmessage{(Inkscape) Color is used for the text in Inkscape, but the package 'color.sty' is not loaded}%
    \renewcommand\color[2][]{}%
  }%
  \providecommand\transparent[1]{%
    \errmessage{(Inkscape) Transparency is used (non-zero) for the text in Inkscape, but the package 'transparent.sty' is not loaded}%
    \renewcommand\transparent[1]{}%
  }%
  \providecommand\rotatebox[2]{#2}%
  \ifx\svgwidth\undefined%
    \setlength{\unitlength}{487.53894424bp}%
    \ifx\svgscale\undefined%
      \relax%
    \else%
      \setlength{\unitlength}{\unitlength * \real{\svgscale}}%
    \fi%
  \else%
    \setlength{\unitlength}{\svgwidth}%
  \fi%
  \global\let\svgwidth\undefined%
  \global\let\svgscale\undefined%
  \makeatother%
  \begin{picture}(1,0.54719888)%
    \put(0,0){\includegraphics[width=\unitlength]{All_exp2.eps}}%
    \put(0.57040093,0.39405695){\color[rgb]{0,0,0}\rotatebox{90}{\makebox(0,0)[b]{\smash{Sample}}}}%
    \put(0.00976101,0.30280086){\color[rgb]{0,0,0}\makebox(0,0)[lt]{\begin{minipage}{0.08637717\unitlength}\centering 10 mm\end{minipage}}}%
    \put(0.00551574,0.10287757){\color[rgb]{0,0,0}\makebox(0,0)[lt]{\begin{minipage}{0.05971561\unitlength}\centering \end{minipage}}}%
    \put(0.40869786,0.47365696){\color[rgb]{0.6745098,0,0}\rotatebox{90}{\makebox(0,0)[b]{\smash{1.0 mm}}}}%
    \put(0.46915115,0.47401185){\color[rgb]{0.6745098,0,0}\rotatebox{90}{\makebox(0,0)[b]{\smash{0.6 mm}}}}%
    \put(0.33997805,0.47401185){\color[rgb]{0.6745098,0,0}\rotatebox{90}{\makebox(0,0)[b]{\smash{0.4 mm}}}}%
    \put(0.34957303,0.29161907){\color[rgb]{0,0,0}\makebox(0,0)[b]{\smash{1 mm}}}%
    \put(0.50530701,0.16463103){\color[rgb]{0,0,0}\makebox(0,0)[b]{\smash{Al}}}%
    \put(0.41306259,0.12407371){\color[rgb]{1,1,1}\makebox(0,0)[b]{\smash{GrAl}}}%
    \put(0.7991529,0.10964438){\color[rgb]{0.23921569,0.45098039,0.99215686}\rotatebox{90}{\makebox(0,0)[b]{\smash{100 mK}}}}%
    \put(0.9096483,0.10998874){\color[rgb]{0.23921569,0.45098039,0.99215686}\rotatebox{90}{\makebox(0,0)[b]{\smash{4 K}}}}%
    \put(0.85337343,0.06134278){\color[rgb]{0,0,0}\makebox(0,0)[b]{\smash{-20 dB}}}%
    \put(0.85134064,0.19703877){\color[rgb]{0,0,0}\makebox(0,0)[b]{\smash{+20 dBm}}}%
    \put(0.34004264,0.05602886){\color[rgb]{1,1,1}\makebox(0,0)[b]{\smash{100 $\mu$m}}}%
    \put(0.87103115,0.48911818){\color[rgb]{0,0,0}\makebox(0,0)[b]{\smash{+43 dBm}}}%
    \put(0.87634568,0.3402953){\color[rgb]{0,0,0}\makebox(0,0)[b]{\smash{-20 dB}}}%
    \put(0.76476616,0.3402953){\color[rgb]{0,0,0}\makebox(0,0)[b]{\smash{-20 dB}}}%
    \put(0.69256765,0.3402953){\color[rgb]{0,0,0}\makebox(0,0)[b]{\smash{-30 dB}}}%
    \put(0.9030848,0.40535023){\color[rgb]{0.23921569,0.45098039,0.99215686}\rotatebox{90}{\makebox(0,0)[b]{\smash{1.6 K}}}}%
    \put(0.72176255,0.40172409){\color[rgb]{0.23921569,0.45098039,0.99215686}\rotatebox{90}{\makebox(0,0)[b]{\smash{20 mK}}}}%
    \put(0.80539614,0.42676429){\color[rgb]{0,0,0}\makebox(0,0)[b]{\smash{12 GHz}}}%
    \put(0.82123781,0.28564718){\color[rgb]{0,0,0}\makebox(0,0)[b]{\smash{12 GHz}}}%
    \put(0.74844302,0.10785779){\color[rgb]{0,0,0}\rotatebox{90}{\makebox(0,0)[b]{\smash{Sample}}}}%
    \put(0.61621538,0.34201955){\color[rgb]{0,0,0}\makebox(0,0)[b]{\smash{8.2 GHz}}}%
    \put(0.7089764,0.00555226){\color[rgb]{0,0,0}\makebox(0,0)[b]{\smash{Cryostat}}}%
    \put(0.57346949,0.27794119){\color[rgb]{0,0,0}\makebox(0,0)[b]{\smash{Cryostat}}}%
    \put(0.96314647,0.10898554){\color[rgb]{0,0,0}\rotatebox{90}{\makebox(0,0)[b]{\smash{Readout}}}}%
    \put(0.98283698,0.10898554){\color[rgb]{0,0,0}\rotatebox{90}{\makebox(0,0)[b]{\smash{electronics}}}}%
    \put(0.57515689,0.10759598){\color[rgb]{0,0,0}\rotatebox{90}{\makebox(0,0)[b]{\smash{Martin-Puplett}}}}%
    \put(0.59156564,0.10759598){\color[rgb]{0,0,0}\rotatebox{90}{\makebox(0,0)[b]{\smash{Interferometer}}}}%
    \put(0.70038233,0.10785779){\color[rgb]{0.4,0.70196078,0.18823529}\rotatebox{90}{\makebox(0,0)[b]{\smash{\textbf{Optics}}}}}%
    \put(0.53076353,0.23199577){\color[rgb]{0,0,0}\makebox(0,0)[b]{\smash{\textbf{High Frequency Setup (<200 GHz): Transmission measurements}}}}%
    \put(0.53076353,0.53063865){\color[rgb]{0,0,0}\makebox(0,0)[b]{\smash{\textbf{Low Frequency Setup (<20 GHz): Reflection measurements}}}}%
    \put(0.03521218,0.49453896){\color[rgb]{0,0,0}\makebox(0,0)[b]{\smash{\textbf{a)}}}}%
    \put(0.02208482,0.18276885){\color[rgb]{0,0,0}\makebox(0,0)[b]{\smash{\textbf{c)}}}}%
    \put(0.55373791,0.49453896){\color[rgb]{0,0,0}\makebox(0,0)[b]{\smash{\textbf{b)}}}}%
    \put(0.56686492,0.19917782){\color[rgb]{0,0,0}\makebox(0,0)[b]{\smash{\textbf{d)}}}}%
    \put(0.00837796,0.04797062){\color[rgb]{1,1,1}\makebox(0,0)[lt]{\begin{minipage}{0.09889983\unitlength}\raggedright 10 mm\end{minipage}}}%
    \put(0.95600295,0.39450098){\color[rgb]{0,0,0}\rotatebox{90}{\makebox(0,0)[b]{\smash{VNA}}}}%
    \put(0.99101799,0.39968233){\color[rgb]{0,0,0}\rotatebox{90}{\makebox(0,0)[b]{\smash{RF Generator}}}}%
  \end{picture}%
\endgroup%

%% file: All_R3.eps_tex
\begingroup%
  \makeatletter%
  \providecommand\color[2][]{%
    \errmessage{(Inkscape) Color is used for the text in Inkscape, but the package 'color.sty' is not loaded}%
    \renewcommand\color[2][]{}%
  }%
  \providecommand\transparent[1]{%
    \errmessage{(Inkscape) Transparency is used (non-zero) for the text in Inkscape, but the package 'transparent.sty' is not loaded}%
    \renewcommand\transparent[1]{}%
  }%
  \providecommand\rotatebox[2]{#2}%
  \ifx\svgwidth\undefined%
    \setlength{\unitlength}{669.21220684bp}%
    \ifx\svgscale\undefined%
      \relax%
    \else%
      \setlength{\unitlength}{\unitlength * \real{\svgscale}}%
    \fi%
  \else%
    \setlength{\unitlength}{\svgwidth}%
  \fi%
  \global\let\svgwidth\undefined%
  \global\let\svgscale\undefined%
  \makeatother%
  \begin{picture}(1,0.56034016)%
    \put(0,0){\includegraphics[width=\unitlength]{All_R3.eps}}%
    \put(0.65855883,0.0389115){\color[rgb]{0,0,0}\makebox(0,0)[b]{\smash{0}}}%
    \put(0.770593,0.0389115){\color[rgb]{0,0,0}\makebox(0,0)[b]{\smash{0}}}%
    \put(0.88246986,0.0389115){\color[rgb]{0,0,0}\makebox(0,0)[b]{\smash{0}}}%
    \put(0.5709829,0.03899306){\color[rgb]{0,0,0}\makebox(0,0)[b]{\smash{25}}}%
    \put(0.62634329,0.03899306){\color[rgb]{0,0,0}\makebox(0,0)[b]{\smash{50}}}%
    \put(0.4986971,0.10865112){\color[rgb]{0,0,0}\makebox(0,0)[b]{\smash{25}}}%
    \put(0.49864089,0.16725349){\color[rgb]{0,0,0}\makebox(0,0)[b]{\smash{50}}}%
    \put(0.4986971,0.2258557){\color[rgb]{0,0,0}\makebox(0,0)[b]{\smash{75}}}%
    \put(0.49059494,0.28445806){\color[rgb]{0,0,0}\makebox(0,0)[b]{\smash{100}}}%
    \put(0.49065115,0.3430588){\color[rgb]{0,0,0}\makebox(0,0)[b]{\smash{125}}}%
    \put(0.49059494,0.40166112){\color[rgb]{0,0,0}\makebox(0,0)[b]{\smash{150}}}%
    \put(0.49065115,0.46026349){\color[rgb]{0,0,0}\makebox(0,0)[b]{\smash{175}}}%
    \put(0.49059494,0.51886585){\color[rgb]{0,0,0}\makebox(0,0)[b]{\smash{200}}}%
    \put(0.57063151,0.22063838){\color[rgb]{0.6745098,0,0}\makebox(0,0)[b]{\smash{\textbf{frequency}}}}%
    \put(0.71185761,0.27588468){\color[rgb]{0.18431373,0.21568627,1}\makebox(0,0)[b]{\smash{\textbf{$2\Delta$ Al}}}}%
    \put(0.93658588,0.53993422){\color[rgb]{0.98039216,0.23529412,0.23529412}\makebox(0,0)[b]{\smash{\textbf{GrAl$\#$3}}}}%
    \put(0.82484079,0.53993422){\color[rgb]{0.42352941,0.65490196,0.07843137}\makebox(0,0)[b]{\smash{\textbf{GrAl$\#$2}}}}%
    \put(0.71309811,0.53993422){\color[rgb]{0.18431373,0.21568627,1}\makebox(0,0)[b]{\smash{\textbf{Al}}}}%
    \put(0.57121771,0.53993422){\color[rgb]{0.6745098,0,0}\makebox(0,0)[b]{\smash{\textbf{GrAl$\#$1}}}}%
    \put(0.09502413,0.20457772){\color[rgb]{0,0,0}\makebox(0,0)[b]{\smash{$10^0$}}}%
    \put(0.19152286,0.20457772){\color[rgb]{0,0,0}\makebox(0,0)[b]{\smash{$10^1$}}}%
    \put(0.28802742,0.20457772){\color[rgb]{0,0,0}\makebox(0,0)[b]{\smash{$10^2$}}}%
    \put(0.38452614,0.20457772){\color[rgb]{0,0,0}\makebox(0,0)[b]{\smash{$10^3$}}}%
    \put(0.03598696,0.21931841){\color[rgb]{0,0,0}\makebox(0,0)[b]{\smash{-100}}}%
    \put(0.04069179,0.23953443){\color[rgb]{0,0,0}\makebox(0,0)[b]{\smash{-80}}}%
    \put(0.04069179,0.25976359){\color[rgb]{0,0,0}\makebox(0,0)[b]{\smash{-60}}}%
    \put(0.04069179,0.27998326){\color[rgb]{0,0,0}\makebox(0,0)[b]{\smash{-40}}}%
    \put(0.04069179,0.30020512){\color[rgb]{0,0,0}\makebox(0,0)[b]{\smash{-20}}}%
    \put(0.04875952,0.32042843){\color[rgb]{0,0,0}\makebox(0,0)[b]{\smash{0}}}%
    \put(0.01203126,0.20225055){\color[rgb]{0,0,0}\rotatebox{90}{\makebox(0,0)[b]{\smash{Fundamental frequency shift, $\Delta f_1$ (kHz)}}}}%
    \put(0.1429653,0.37020215){\color[rgb]{0,0,0}\makebox(0,0)[b]{\smash{Frequency, $f$ (GHz)}}}%
    \put(0.14352682,0.54018521){\color[rgb]{0,0,0}\makebox(0,0)[b]{\smash{$| S_{11}|$, (dB)}}}%
    \put(0.37047328,0.37020215){\color[rgb]{0,0,0}\makebox(0,0)[b]{\smash{Frequency, $f$ (GHz)}}}%
    \put(0.37063563,0.5402249){\color[rgb]{0,0,0}\makebox(0,0)[b]{\smash{arg$(S_{11})$, (rad)}}}%
    \put(0.06801009,0.03813384){\color[rgb]{0,0,0}\makebox(0,0)[b]{\smash{0}}}%
    \put(0.16320219,0.03813384){\color[rgb]{0,0,0}\makebox(0,0)[b]{\smash{1000}}}%
    \put(0.26003538,0.03813019){\color[rgb]{0,0,0}\makebox(0,0)[b]{\smash{2000}}}%
    \put(0.45101188,0.03813384){\color[rgb]{0,0,0}\makebox(0,0)[b]{\smash{4000}}}%
    \put(0.24664991,0.01250355){\color[rgb]{0,0,0}\makebox(0,0)[b]{\smash{Average photon number, $\bar{N}$}}}%
    \put(0.71258633,0.50806832){\color[rgb]{0,0,0}\makebox(0,0)[b]{\smash{180 GHz}}}%
    \put(0.03517648,0.41830781){\color[rgb]{0,0,0}\makebox(0,0)[b]{\smash{-0.10}}}%
    \put(0.03517648,0.45197022){\color[rgb]{0,0,0}\makebox(0,0)[b]{\smash{-0.05}}}%
    \put(0.03906179,0.4856327){\color[rgb]{0,0,0}\makebox(0,0)[b]{\smash{0.00}}}%
    \put(0.0843176,0.39396701){\color[rgb]{0,0,0}\makebox(0,0)[b]{\smash{6.285}}}%
    \put(0.14067935,0.39403476){\color[rgb]{0,0,0}\makebox(0,0)[b]{\smash{6.278}}}%
    \put(0.19717666,0.39401834){\color[rgb]{0,0,0}\makebox(0,0)[b]{\smash{6.289}}}%
    \put(0.26497457,0.42670322){\color[rgb]{0,0,0}\makebox(0,0)[b]{\smash{-2.5}}}%
    \put(0.26804619,0.46800853){\color[rgb]{0,0,0}\makebox(0,0)[b]{\smash{0.0}}}%
    \put(0.26804619,0.50930074){\color[rgb]{0,0,0}\makebox(0,0)[b]{\smash{2.5}}}%
    \put(0.31015268,0.39396701){\color[rgb]{0,0,0}\makebox(0,0)[b]{\smash{6.285}}}%
    \put(0.37002244,0.39403476){\color[rgb]{0,0,0}\makebox(0,0)[b]{\smash{6.278}}}%
    \put(0.43002237,0.39401834){\color[rgb]{0,0,0}\makebox(0,0)[b]{\smash{6.289}}}%
    \put(0.03598696,0.05316514){\color[rgb]{0,0,0}\makebox(0,0)[b]{\smash{-100}}}%
    \put(0.04069179,0.07425235){\color[rgb]{0,0,0}\makebox(0,0)[b]{\smash{-80}}}%
    \put(0.04069179,0.09528571){\color[rgb]{0,0,0}\makebox(0,0)[b]{\smash{-60}}}%
    \put(0.04069179,0.11630936){\color[rgb]{0,0,0}\makebox(0,0)[b]{\smash{-40}}}%
    \put(0.04069179,0.1373355){\color[rgb]{0,0,0}\makebox(0,0)[b]{\smash{-20}}}%
    \put(0.04875952,0.15784921){\color[rgb]{0,0,0}\makebox(0,0)[b]{\smash{0}}}%
    \put(0.35644676,0.03813019){\color[rgb]{0,0,0}\makebox(0,0)[b]{\smash{3000}}}%
    \put(0.57027889,0.01250355){\color[rgb]{0,0,0}\makebox(0,0)[b]{\smash{Mode number}}}%
    \put(0.82393696,0.01250355){\color[rgb]{0,0,0}\makebox(0,0)[b]{\smash{MPI response (Hz)}}}%
    \put(0.44440364,0.4279494){\color[rgb]{0,0,0}\makebox(0,0)[rb]{\smash{$Q_c = 10^4$}}}%
    \put(0.21800266,0.4279494){\color[rgb]{0,0,0}\makebox(0,0)[rb]{\smash{$Q_i = 10^5$}}}%
    \put(0.47346934,0.29603714){\color[rgb]{0,0,0}\rotatebox{90}{\makebox(0,0)[b]{\smash{Frequency (GHz)}}}}%
    \put(0.03805403,0.53993422){\color[rgb]{0,0,0}\makebox(0,0)[b]{\smash{\textbf{a)}}}}%
    \put(0.25801393,0.53993422){\color[rgb]{0,0,0}\makebox(0,0)[b]{\smash{\textbf{b)}}}}%
    \put(0.03805403,0.34746914){\color[rgb]{0,0,0}\makebox(0,0)[b]{\smash{\textbf{c)}}}}%
    \put(0.03805403,0.18608519){\color[rgb]{0,0,0}\makebox(0,0)[b]{\smash{\textbf{d)}}}}%
    \put(0.48036485,0.53993422){\color[rgb]{0,0,0}\makebox(0,0)[b]{\smash{\textbf{e)}}}}%
    \put(0.64533478,0.53993422){\color[rgb]{0,0,0}\makebox(0,0)[b]{\smash{\textbf{f)}}}}%
    \put(0.09944966,0.24962435){\color[rgb]{0,0,0}\makebox(0,0)[lb]{\smash{$K_{11} = 21$ Hz}}}%
    \put(0.57063151,0.23498361){\color[rgb]{0.6745098,0,0}\makebox(0,0)[b]{\smash{\textbf{Plasma}}}}%
    \put(0.82445131,0.39199908){\color[rgb]{0.42352941,0.65490196,0.07843137}\makebox(0,0)[b]{\smash{\textbf{$2\Delta$}}}}%
    \put(0.93693596,0.42367815){\color[rgb]{0.98039216,0.23529412,0.23529412}\makebox(0,0)[b]{\smash{\textbf{$2\Delta$}}}}%
    \put(0.74947597,0.0389115){\color[rgb]{0,0,0}\makebox(0,0)[b]{\smash{50}}}%
    \put(0.85668331,0.0389115){\color[rgb]{0,0,0}\makebox(0,0)[b]{\smash{100}}}%
    \put(0.97630503,0.0389115){\color[rgb]{0,0,0}\makebox(0,0)[b]{\smash{2}}}%
    \put(0.59062202,0.13557131){\color[rgb]{0,0,0}\rotatebox{90}{\makebox(0,0)[b]{\smash{$K_{1n}$ (Hz)}}}}%
    \put(0.60923704,0.10460567){\color[rgb]{0,0,0}\makebox(0,0)[b]{\smash{10}}}%
    \put(0.60923704,0.15564689){\color[rgb]{0,0,0}\makebox(0,0)[b]{\smash{20}}}%
    \put(0.82445131,0.37526297){\color[rgb]{0.42352941,0.65490196,0.07843137}\makebox(0,0)[b]{\smash{\textbf{GrAl$\#$2}}}}%
    \put(0.93693596,0.40694204){\color[rgb]{0.98039216,0.23529412,0.23529412}\makebox(0,0)[b]{\smash{\textbf{GrAl$\#$3}}}}%
    \put(0.09944966,0.28070567){\color[rgb]{0,0,0}\makebox(0,0)[lb]{\smash{$\Delta f_1 = -K_{11} \bar{N}$}}}%
  \end{picture}%
\endgroup%

%% file: K11_3.eps_tex
\begingroup%
  \makeatletter%
  \providecommand\color[2][]{%
    \errmessage{(Inkscape) Color is used for the text in Inkscape, but the package 'color.sty' is not loaded}%
    \renewcommand\color[2][]{}%
  }%
  \providecommand\transparent[1]{%
    \errmessage{(Inkscape) Transparency is used (non-zero) for the text in Inkscape, but the package 'transparent.sty' is not loaded}%
    \renewcommand\transparent[1]{}%
  }%
  \providecommand\rotatebox[2]{#2}%
  \ifx\svgwidth\undefined%
    \setlength{\unitlength}{1462.33007813bp}%
    \ifx\svgscale\undefined%
      \relax%
    \else%
      \setlength{\unitlength}{\unitlength * \real{\svgscale}}%
    \fi%
  \else%
    \setlength{\unitlength}{\svgwidth}%
  \fi%
  \global\let\svgwidth\undefined%
  \global\let\svgscale\undefined%
  \makeatother%
  \begin{picture}(1,0.43885652)%
    \put(0,0){\includegraphics[width=\unitlength]{K11_3.eps}}%
    \put(0.06047644,0.03610985){\color[rgb]{0,0,0}\makebox(0,0)[b]{\smash{$10^{-3}$}}}%
    \put(0.12034823,0.03610985){\color[rgb]{0,0,0}\makebox(0,0)[b]{\smash{$10^{-2}$}}}%
    \put(0.18004547,0.03610985){\color[rgb]{0,0,0}\makebox(0,0)[b]{\smash{$10^{-1}$}}}%
    \put(0.23974254,0.03610985){\color[rgb]{0,0,0}\makebox(0,0)[b]{\smash{1}}}%
    \put(0.29979007,0.03610985){\color[rgb]{0,0,0}\makebox(0,0)[b]{\smash{$10^{1}$}}}%
    \put(0.35966239,0.03610985){\color[rgb]{0,0,0}\makebox(0,0)[b]{\smash{$10^2$}}}%
    \put(0.41935963,0.03610985){\color[rgb]{0,0,0}\makebox(0,0)[b]{\smash{$10^3$}}}%
    \put(0.47923193,0.03610985){\color[rgb]{0,0,0}\makebox(0,0)[b]{\smash{$10^4$}}}%
    \put(0.53910347,0.03610985){\color[rgb]{0,0,0}\makebox(0,0)[b]{\smash{$10^5$}}}%
    \put(0.03658254,0.0694978){\color[rgb]{0,0,0}\makebox(0,0)[b]{\smash{$10^{-2}$}}}%
    \put(0.03658254,0.11974146){\color[rgb]{0,0,0}\makebox(0,0)[b]{\smash{$10^{-1}$}}}%
    \put(0.03640731,0.1693925){\color[rgb]{0,0,0}\makebox(0,0)[b]{\smash{1}}}%
    \put(0.03658254,0.21958298){\color[rgb]{0,0,0}\makebox(0,0)[b]{\smash{$10^1$}}}%
    \put(0.03658254,0.26982673){\color[rgb]{0,0,0}\makebox(0,0)[b]{\smash{$10^2$}}}%
    \put(0.03658254,0.32007046){\color[rgb]{0,0,0}\makebox(0,0)[b]{\smash{$10^3$}}}%
    \put(0.03658254,0.3701384){\color[rgb]{0,0,0}\makebox(0,0)[b]{\smash{$10^4$}}}%
    \put(0.03658254,0.42038213){\color[rgb]{0,0,0}\makebox(0,0)[b]{\smash{$10^5$}}}%
    \put(0.00907316,0.24011332){\color[rgb]{0,0,0}\rotatebox{90}{\makebox(0,0)[b]{\smash{Self-Kerr, $K_{11}/2 \pi$ (Hz)}}}}%
    \put(0.29978996,0.00897479){\color[rgb]{0,0,0}\makebox(0,0)[b]{\smash{$f_1^2\ j_c^{-1} V_{GrAl}^{-1}$ (GHz$^2$mA$^{-1}\upmu$m$^{-1}$)}}}%
    \put(0.66694111,0.4214651){\color[rgb]{0,0,0}\makebox(0,0)[b]{\smash{$\rho$, $\upmu\Omega\,$cm}}}%
    \put(0.66694111,0.38931903){\color[rgb]{0,0,0}\makebox(0,0)[b]{\smash{2000}}}%
    \put(0.66694111,0.36445084){\color[rgb]{0,0,0}\makebox(0,0)[b]{\smash{2000}}}%
    \put(0.66694111,0.33958642){\color[rgb]{0,0,0}\makebox(0,0)[b]{\smash{4000}}}%
    \put(0.66694111,0.31471825){\color[rgb]{0,0,0}\makebox(0,0)[b]{\smash{4000}}}%
    \put(0.66694111,0.28984644){\color[rgb]{0,0,0}\makebox(0,0)[b]{\smash{2800}}}%
    \put(0.66694111,0.26497832){\color[rgb]{0,0,0}\makebox(0,0)[b]{\smash{2800}}}%
    \put(0.66694111,0.24011385){\color[rgb]{0,0,0}\makebox(0,0)[b]{\smash{4000}}}%
    \put(0.66694111,0.21524204){\color[rgb]{0,0,0}\makebox(0,0)[b]{\smash{2800}}}%
    \put(0.66694111,0.19037757){\color[rgb]{0,0,0}\makebox(0,0)[b]{\smash{1600}}}%
    \put(0.66694111,0.16550944){\color[rgb]{0,0,0}\makebox(0,0)[b]{\smash{900}}}%
    \put(0.66694111,0.14064131){\color[rgb]{0,0,0}\makebox(0,0)[b]{\smash{40}}}%
    \put(0.66694111,0.1157695){\color[rgb]{0,0,0}\makebox(0,0)[b]{\smash{80}}}%
    \put(0.66694111,0.09090136){\color[rgb]{0,0,0}\makebox(0,0)[b]{\smash{220}}}%
    \put(0.66694111,0.0660369){\color[rgb]{0,0,0}\makebox(0,0)[b]{\smash{160}}}%
    \put(0.77711599,0.4214651){\color[rgb]{0,0,0}\makebox(0,0)[b]{\smash{$V_{GrAl}$, $\upmu$m$^3$}}}%
    \put(0.77711599,0.38931903){\color[rgb]{0,0,0}\makebox(0,0)[b]{\smash{0.003}}}%
    \put(0.77711599,0.36445451){\color[rgb]{0,0,0}\makebox(0,0)[b]{\smash{0.01}}}%
    \put(0.77711599,0.33958274){\color[rgb]{0,0,0}\makebox(0,0)[b]{\smash{43.2}}}%
    \put(0.77711599,0.31471458){\color[rgb]{0,0,0}\makebox(0,0)[b]{\smash{120}}}%
    \put(0.77711599,0.28985011){\color[rgb]{0,0,0}\makebox(0,0)[b]{\smash{107}}}%
    \put(0.77711599,0.26497832){\color[rgb]{0,0,0}\makebox(0,0)[b]{\smash{88}}}%
    \put(0.77711599,0.24011018){\color[rgb]{0,0,0}\makebox(0,0)[b]{\smash{800}}}%
    \put(0.77711599,0.21524204){\color[rgb]{0,0,0}\makebox(0,0)[b]{\smash{624}}}%
    \put(0.77711599,0.1903739){\color[rgb]{0,0,0}\makebox(0,0)[b]{\smash{288}}}%
    \put(0.77711599,0.16550577){\color[rgb]{0,0,0}\makebox(0,0)[b]{\smash{288}}}%
    \put(0.77711599,0.14064131){\color[rgb]{0,0,0}\makebox(0,0)[b]{\smash{100}}}%
    \put(0.77711599,0.11577317){\color[rgb]{0,0,0}\makebox(0,0)[b]{\smash{100}}}%
    \put(0.77711599,0.09090504){\color[rgb]{0,0,0}\makebox(0,0)[b]{\smash{100}}}%
    \put(0.77711599,0.0660369){\color[rgb]{0,0,0}\makebox(0,0)[b]{\smash{100}}}%
    \put(0.88722672,0.4214651){\color[rgb]{0,0,0}\makebox(0,0)[b]{\smash{$f_1$, GHz}}}%
    \put(0.88722672,0.38931903){\color[rgb]{0,0,0}\makebox(0,0)[b]{\smash{4.7}}}%
    \put(0.88722672,0.3645647){\color[rgb]{0,0,0}\makebox(0,0)[b]{\smash{5}}}%
    \put(0.88722672,0.33958274){\color[rgb]{0,0,0}\makebox(0,0)[b]{\smash{7}}}%
    \put(0.88722672,0.31471458){\color[rgb]{0,0,0}\makebox(0,0)[b]{\smash{6.3}}}%
    \put(0.88722672,0.28985746){\color[rgb]{0,0,0}\makebox(0,0)[b]{\smash{7.6}}}%
    \put(0.88722672,0.26486446){\color[rgb]{0,0,0}\makebox(0,0)[b]{\smash{7.2}}}%
    \put(0.88722672,0.2401212){\color[rgb]{0,0,0}\makebox(0,0)[b]{\smash{6}}}%
    \put(0.88722672,0.21524204){\color[rgb]{0,0,0}\makebox(0,0)[b]{\smash{8.6}}}%
    \put(0.88722672,0.1903739){\color[rgb]{0,0,0}\makebox(0,0)[b]{\smash{3.2}}}%
    \put(0.88722672,0.16550577){\color[rgb]{0,0,0}\makebox(0,0)[b]{\smash{3.5}}}%
    \put(0.88722672,0.14063764){\color[rgb]{0,0,0}\makebox(0,0)[b]{\smash{5.2}}}%
    \put(0.88722672,0.11575113){\color[rgb]{0,0,0}\makebox(0,0)[b]{\smash{4.1}}}%
    \put(0.88722672,0.09090136){\color[rgb]{0,0,0}\makebox(0,0)[b]{\smash{2.6}}}%
    \put(0.88722672,0.06603323){\color[rgb]{0,0,0}\makebox(0,0)[b]{\smash{2.6}}}%
    \put(0.33125764,0.11707879){\color[rgb]{0,0,0}\makebox(0,0)[lb]{\smash{$K_{11}=\frac{3 }{16} \pi e a \frac{ \omega_1^2}{ j_c V_{GrAl}}$}}}%
    \put(0.95967273,0.31572097){\color[rgb]{0,0,0}\makebox(0,0)[b]{\smash{GrAl$\#$1}}}%
    \put(0.95967273,0.11677589){\color[rgb]{0,0,0}\makebox(0,0)[b]{\smash{GrAl$\#$2}}}%
  \end{picture}%
\endgroup%

%% file: CD.eps_tex
\begingroup%
  \makeatletter%
  \providecommand\color[2][]{%
    \errmessage{(Inkscape) Color is used for the text in Inkscape, but the package 'color.sty' is not loaded}%
    \renewcommand\color[2][]{}%
  }%
  \providecommand\transparent[1]{%
    \errmessage{(Inkscape) Transparency is used (non-zero) for the text in Inkscape, but the package 'transparent.sty' is not loaded}%
    \renewcommand\transparent[1]{}%
  }%
  \providecommand\rotatebox[2]{#2}%
  \ifx\svgwidth\undefined%
    \setlength{\unitlength}{469.41768265bp}%
    \ifx\svgscale\undefined%
      \relax%
    \else%
      \setlength{\unitlength}{\unitlength * \real{\svgscale}}%
    \fi%
  \else%
    \setlength{\unitlength}{\svgwidth}%
  \fi%
  \global\let\svgwidth\undefined%
  \global\let\svgscale\undefined%
  \makeatother%
  \begin{picture}(1,0.59558385)%
    \put(0,0){\includegraphics[width=\unitlength]{CD.eps}}%
    \put(0.92690979,0.05351308){\color[rgb]{0,0,0}\makebox(0,0)[b]{\smash{$10^1$}}}%
    \put(0.92690979,0.15564972){\color[rgb]{0,0,0}\makebox(0,0)[b]{\smash{$10^1$}}}%
    \put(0.92690979,0.25778636){\color[rgb]{0,0,0}\makebox(0,0)[b]{\smash{$10^2$}}}%
    \put(0.92690979,0.35992298){\color[rgb]{0,0,0}\makebox(0,0)[b]{\smash{$10^3$}}}%
    \put(0.9269099,0.46205963){\color[rgb]{0,0,0}\makebox(0,0)[b]{\smash{$10^4$}}}%
    \put(0.09708736,0.33107539){\color[rgb]{0,0,0}\makebox(0,0)[b]{\smash{$0$}}}%
    \put(0.09708736,0.52636707){\color[rgb]{0,0,0}\makebox(0,0)[b]{\smash{$1$}}}%
    \put(0.13938609,0.03039415){\color[rgb]{0,0,0}\makebox(0,0)[b]{\smash{$0$}}}%
    \put(0.17817893,0.03039415){\color[rgb]{0,0,0}\makebox(0,0)[b]{\smash{$1$}}}%
    \put(0.09671289,0.06204356){\color[rgb]{0,0,0}\makebox(0,0)[b]{\smash{$0$}}}%
    \put(0.09671289,0.2573352){\color[rgb]{0,0,0}\makebox(0,0)[b]{\smash{$1$}}}%
    \put(0.02618816,0.42708437){\color[rgb]{0,0,0}\makebox(0,0)[b]{\smash{$J_y$}}}%
    \put(0.02618816,0.16128677){\color[rgb]{0,0,0}\makebox(0,0)[b]{\smash{$J_x$}}}%
    \put(0.29763754,0.03039415){\color[rgb]{0,0,0}\makebox(0,0)[b]{\smash{$0$}}}%
    \put(0.33643039,0.03039415){\color[rgb]{0,0,0}\makebox(0,0)[b]{\smash{$1$}}}%
    \put(0.45588893,0.03039415){\color[rgb]{0,0,0}\makebox(0,0)[b]{\smash{$0$}}}%
    \put(0.49468179,0.03039415){\color[rgb]{0,0,0}\makebox(0,0)[b]{\smash{$1$}}}%
    \put(0.61414043,0.03039415){\color[rgb]{0,0,0}\makebox(0,0)[b]{\smash{$0$}}}%
    \put(0.65293328,0.03039415){\color[rgb]{0,0,0}\makebox(0,0)[b]{\smash{$1$}}}%
    \put(0.77205663,0.03039415){\color[rgb]{0,0,0}\makebox(0,0)[b]{\smash{$0$}}}%
    \put(0.81084948,0.03039415){\color[rgb]{0,0,0}\makebox(0,0)[b]{\smash{$1$}}}%
    \put(0.09442449,0.43479058){\color[rgb]{0,0,0}\rotatebox{90}{\makebox(0,0)[b]{\smash{$y / \ell$}}}}%
    \put(0.15668463,0.00085208){\color[rgb]{0,0,0}\makebox(0,0)[b]{\smash{$x / b$}}}%
    \put(0.09442449,0.1662963){\color[rgb]{0,0,0}\rotatebox{90}{\makebox(0,0)[b]{\smash{$y / \ell$}}}}%
    \put(0.31493607,0.00085208){\color[rgb]{0,0,0}\makebox(0,0)[b]{\smash{$x / b$}}}%
    \put(0.47318747,0.00085208){\color[rgb]{0,0,0}\makebox(0,0)[b]{\smash{$x / b$}}}%
    \put(0.63143896,0.00085208){\color[rgb]{0,0,0}\makebox(0,0)[b]{\smash{$x / b$}}}%
    \put(0.79276365,0.00085208){\color[rgb]{0,0,0}\makebox(0,0)[b]{\smash{$x / b$}}}%
    \put(0.15668467,0.58193525){\color[rgb]{0,0,0}\makebox(0,0)[b]{\smash{Mode 1}}}%
    \put(0.31493607,0.5814318){\color[rgb]{0,0,0}\makebox(0,0)[b]{\smash{Mode 2}}}%
    \put(0.47318747,0.57804249){\color[rgb]{0,0,0}\makebox(0,0)[b]{\smash{Mode 3}}}%
    \put(0.63143896,0.57917226){\color[rgb]{0,0,0}\makebox(0,0)[b]{\smash{Mode 4}}}%
    \put(0.78969036,0.57917226){\color[rgb]{0,0,0}\makebox(0,0)[b]{\smash{Mode 5}}}%
    \put(0.99716864,0.29877846){\color[rgb]{0,0,0}\rotatebox{90}{\makebox(0,0)[b]{\smash{$J$ (A/m) }}}}%
  \end{picture}%
\endgroup%

%% file: samples.eps_tex
\begingroup%
  \makeatletter%
  \providecommand\color[2][]{%
    \errmessage{(Inkscape) Color is used for the text in Inkscape, but the package 'color.sty' is not loaded}%
    \renewcommand\color[2][]{}%
  }%
  \providecommand\transparent[1]{%
    \errmessage{(Inkscape) Transparency is used (non-zero) for the text in Inkscape, but the package 'transparent.sty' is not loaded}%
    \renewcommand\transparent[1]{}%
  }%
  \providecommand\rotatebox[2]{#2}%
  \ifx\svgwidth\undefined%
    \setlength{\unitlength}{464.97874428bp}%
    \ifx\svgscale\undefined%
      \relax%
    \else%
      \setlength{\unitlength}{\unitlength * \real{\svgscale}}%
    \fi%
  \else%
    \setlength{\unitlength}{\svgwidth}%
  \fi%
  \global\let\svgwidth\undefined%
  \global\let\svgscale\undefined%
  \makeatother%
  \begin{picture}(1,0.86566052)%
    \put(0,0){\includegraphics[width=\unitlength]{samples.eps}}%
    \put(0.89451383,0.13595726){\color[rgb]{0.66666667,0,0}\makebox(0,0)[b]{\smash{\textbf{GrAl}}}}%
    \put(0.06151639,0.03339566){\color[rgb]{1,1,1}\makebox(0,0)[b]{\smash{\textbf{Al}}}}%
    \put(0.40874328,0.03339566){\color[rgb]{1,1,1}\makebox(0,0)[b]{\smash{\textbf{Al}}}}%
    \put(0.74009486,0.03339566){\color[rgb]{1,1,1}\makebox(0,0)[b]{\smash{\textbf{Al}}}}%
    \put(0.6699817,0.6808854){\color[rgb]{0.78039216,0.78039216,0.78039216}\makebox(0,0)[b]{\smash{\textbf{Al}}}}%
    \put(0.66923385,0.77513307){\color[rgb]{0,0,0.66666667}\makebox(0,0)[b]{\smash{\textbf{GrAl}}}}%
    \put(0.74290938,0.5088012){\color[rgb]{0,0,0}\makebox(0,0)[b]{\smash{1}}}%
    \put(0.85971662,0.28971167){\color[rgb]{0,0,0}\makebox(0,0)[b]{\smash{1}}}%
    \put(0.81120824,0.28977781){\color[rgb]{0,0,0}\makebox(0,0)[b]{\smash{0.5}}}%
    \put(0.7208694,0.3606842){\color[rgb]{0,0,0}\makebox(0,0)[b]{\smash{0.25}}}%
    \put(0.7208694,0.45947293){\color[rgb]{0,0,0}\makebox(0,0)[b]{\smash{0.75}}}%
    \put(0.76213903,0.54308818){\color[rgb]{0,0,0}\makebox(0,0)[b]{\smash{$x/\ell$}}}%
    \put(0.55258727,0.44717226){\color[rgb]{0.28235294,0.56078431,0.31764706}\makebox(0,0)[b]{\smash{\textbf{GrAl}}}}%
    \put(0.02700844,0.83444688){\color[rgb]{0,0,0}\makebox(0,0)[b]{\smash{\textbf{a)}}}}%
    \put(0.35987156,0.83156655){\color[rgb]{0,0,0}\makebox(0,0)[b]{\smash{\textbf{b)}}}}%
    \put(0.71271146,0.83103018){\color[rgb]{0,0,0}\makebox(0,0)[b]{\smash{\textbf{c)}}}}%
    \put(0.02630277,0.56455003){\color[rgb]{0,0,0}\makebox(0,0)[b]{\smash{\textbf{d)}}}}%
    \put(0.36028318,0.56455003){\color[rgb]{0,0,0}\makebox(0,0)[b]{\smash{\textbf{e)}}}}%
    \put(0.71426981,0.56455003){\color[rgb]{0,0,0}\makebox(0,0)[b]{\smash{\textbf{f)}}}}%
    \put(0.02703785,0.21918926){\color[rgb]{0,0,0}\makebox(0,0)[b]{\smash{\textbf{g)}}}}%
    \put(0.35987156,0.21904229){\color[rgb]{0,0,0}\makebox(0,0)[b]{\smash{\textbf{h)}}}}%
    \put(0.71562235,0.21904229){\color[rgb]{0,0,0}\makebox(0,0)[b]{\smash{\textbf{i)}}}}%
    \put(0.49429321,0.83075753){\color[rgb]{0,0,0}\makebox(0,0)[b]{\smash{400 $\upmu$m}}}%
    \put(0.82536461,0.83075753){\color[rgb]{0,0,0}\makebox(0,0)[b]{\smash{400 $\upmu$m}}}%
    \put(0.57930919,0.02905265){\color[rgb]{1,1,1}\makebox(0,0)[b]{\smash{10 $\upmu$m}}}%
    \put(0.75054514,0.12253623){\color[rgb]{1,1,1}\makebox(0,0)[b]{\smash{0.19 $\upmu$m}}}%
    \put(0.89703872,0.09188511){\color[rgb]{1,1,1}\makebox(0,0)[b]{\smash{1.5 $\upmu$m}}}%
    \put(0.2628216,0.02765769){\color[rgb]{1,1,1}\makebox(0,0)[b]{\smash{100 $\upmu$m}}}%
    \put(0.5800469,0.34376248){\color[rgb]{1,1,1}\makebox(0,0)[b]{\smash{200 $\upmu$m}}}%
    \put(0.38092235,0.46618158){\color[rgb]{1,1,1}\makebox(0,0)[b]{\smash{x}}}%
    \put(0.42851809,0.51339602){\color[rgb]{1,1,1}\makebox(0,0)[b]{\smash{y}}}%
    \put(0.92316061,0.2883195){\color[rgb]{0,0,0}\makebox(0,0)[b]{\smash{$I(x) / I_0$}}}%
    \put(0.72908552,0.41007852){\color[rgb]{0,0,0}\makebox(0,0)[b]{\smash{0.5}}}%
    \put(0.26451725,0.64555304){\color[rgb]{1,1,1}\makebox(0,0)[b]{\smash{10 mm}}}%
    \put(0.26410077,0.31865686){\color[rgb]{1,1,1}\makebox(0,0)[b]{\smash{10 mm}}}%
  \end{picture}%
\endgroup%

%% file: Interferogram3.eps_tex
\begingroup%
  \makeatletter%
  \providecommand\color[2][]{%
    \errmessage{(Inkscape) Color is used for the text in Inkscape, but the package 'color.sty' is not loaded}%
    \renewcommand\color[2][]{}%
  }%
  \providecommand\transparent[1]{%
    \errmessage{(Inkscape) Transparency is used (non-zero) for the text in Inkscape, but the package 'transparent.sty' is not loaded}%
    \renewcommand\transparent[1]{}%
  }%
  \providecommand\rotatebox[2]{#2}%
  \ifx\svgwidth\undefined%
    \setlength{\unitlength}{686.40004146bp}%
    \ifx\svgscale\undefined%
      \relax%
    \else%
      \setlength{\unitlength}{\unitlength * \real{\svgscale}}%
    \fi%
  \else%
    \setlength{\unitlength}{\svgwidth}%
  \fi%
  \global\let\svgwidth\undefined%
  \global\let\svgscale\undefined%
  \makeatother%
  \begin{picture}(1,0.3962704)%
    \put(0,0){\includegraphics[width=\unitlength]{Interferogram3.eps}}%
    \put(0.79336459,0.00281359){\color[rgb]{0,0,0}\makebox(0,0)[b]{\smash{Frequency (GHz)}}}%
    \put(0.59364589,0.02521427){\color[rgb]{0,0,0}\makebox(0,0)[b]{\smash{0}}}%
    \put(0.72559116,0.02521427){\color[rgb]{0,0,0}\makebox(0,0)[b]{\smash{100}}}%
    \put(0.86069422,0.02521072){\color[rgb]{0,0,0}\makebox(0,0)[b]{\smash{200}}}%
    \put(0.98440399,0.02521072){\color[rgb]{0,0,0}\makebox(0,0)[b]{\smash{300}}}%
    \put(0.28422011,0.00281359){\color[rgb]{0,0,0}\makebox(0,0)[b]{\smash{$dx$ (mm)}}}%
    \put(0.28427816,0.02521427){\color[rgb]{0,0,0}\makebox(0,0)[b]{\smash{0}}}%
    \put(0.09233561,0.02521427){\color[rgb]{0,0,0}\makebox(0,0)[b]{\smash{-10}}}%
    \put(0.3854469,0.02521072){\color[rgb]{0,0,0}\makebox(0,0)[b]{\smash{5}}}%
    \put(0.47492886,0.02521072){\color[rgb]{0,0,0}\makebox(0,0)[b]{\smash{10}}}%
    \put(0.18302747,0.02521427){\color[rgb]{0,0,0}\makebox(0,0)[b]{\smash{-5}}}%
    \put(0.01069438,0.21744475){\color[rgb]{0,0,0}\rotatebox{90}{\makebox(0,0)[b]{\smash{Fundamental frequency shift (Hz)}}}}%
    \put(0.06516398,0.08604746){\color[rgb]{0,0,0}\makebox(0,0)[b]{\smash{0}}}%
    \put(0.0563651,0.129727){\color[rgb]{0,0,0}\makebox(0,0)[b]{\smash{300}}}%
    \put(0.05318274,0.04114893){\color[rgb]{0,0,0}\makebox(0,0)[b]{\smash{-300}}}%
    \put(0.06516398,0.33872521){\color[rgb]{0,0,0}\makebox(0,0)[b]{\smash{0}}}%
    \put(0.05191253,0.38240475){\color[rgb]{0,0,0}\makebox(0,0)[b]{\smash{3000}}}%
    \put(0.04873016,0.29382668){\color[rgb]{0,0,0}\makebox(0,0)[b]{\smash{-3000}}}%
    \put(0.06516398,0.21226836){\color[rgb]{0,0,0}\makebox(0,0)[b]{\smash{0}}}%
    \put(0.0517031,0.25642529){\color[rgb]{0,0,0}\makebox(0,0)[b]{\smash{10000}}}%
    \put(0.04852074,0.16866237){\color[rgb]{0,0,0}\makebox(0,0)[b]{\smash{-10000}}}%
    \put(0.53750083,0.21744475){\color[rgb]{0,0,0}\rotatebox{90}{\makebox(0,0)[b]{\smash{MPI response (Hz)}}}}%
    \put(0.57535527,0.08028355){\color[rgb]{0,0,0}\makebox(0,0)[b]{\smash{1}}}%
    \put(0.56588329,0.24966533){\color[rgb]{0,0,0}\makebox(0,0)[b]{\smash{100}}}%
    \put(0.57033587,0.21124257){\color[rgb]{0,0,0}\makebox(0,0)[b]{\smash{50}}}%
    \put(0.57033587,0.3757379){\color[rgb]{0,0,0}\makebox(0,0)[b]{\smash{50}}}%
    \put(0.57040416,0.33731877){\color[rgb]{0,0,0}\makebox(0,0)[b]{\smash{25}}}%
    \put(0.14093341,0.11675081){\color[rgb]{0.98039216,0.23529412,0.23529412}\makebox(0,0)[b]{\smash{\textbf{GrAl$\#$3}}}}%
    \put(0.14093341,0.24153879){\color[rgb]{0.42352941,0.65490196,0.07843137}\makebox(0,0)[b]{\smash{\textbf{GrAl$\#$2}}}}%
    \put(0.11225687,0.36768739){\color[rgb]{0.18431373,0.21568627,1}\makebox(0,0)[b]{\smash{\textbf{Al}}}}%
    \put(0.63743687,0.11675081){\color[rgb]{0.98039216,0.23529412,0.23529412}\makebox(0,0)[b]{\smash{\textbf{GrAl$\#$3}}}}%
    \put(0.63743687,0.24153879){\color[rgb]{0.42352941,0.65490196,0.07843137}\makebox(0,0)[b]{\smash{\textbf{GrAl$\#$2}}}}%
    \put(0.60876031,0.36768739){\color[rgb]{0.18431373,0.21568627,1}\makebox(0,0)[b]{\smash{\textbf{Al}}}}%
    \put(0.57477479,0.12155057){\color[rgb]{0,0,0}\makebox(0,0)[b]{\smash{2}}}%
  \end{picture}%
\endgroup%

%% file: Ic1.eps_tex
\begingroup%
  \makeatletter%
  \providecommand\color[2][]{%
    \errmessage{(Inkscape) Color is used for the text in Inkscape, but the package 'color.sty' is not loaded}%
    \renewcommand\color[2][]{}%
  }%
  \providecommand\transparent[1]{%
    \errmessage{(Inkscape) Transparency is used (non-zero) for the text in Inkscape, but the package 'transparent.sty' is not loaded}%
    \renewcommand\transparent[1]{}%
  }%
  \providecommand\rotatebox[2]{#2}%
  \ifx\svgwidth\undefined%
    \setlength{\unitlength}{584.92132232bp}%
    \ifx\svgscale\undefined%
      \relax%
    \else%
      \setlength{\unitlength}{\unitlength * \real{\svgscale}}%
    \fi%
  \else%
    \setlength{\unitlength}{\svgwidth}%
  \fi%
  \global\let\svgwidth\undefined%
  \global\let\svgscale\undefined%
  \makeatother%
  \begin{picture}(1,0.4753001)%
    \put(0,0){\includegraphics[width=\unitlength]{Ic1.eps}}%
    \put(0.11219456,0.04784434){\color[rgb]{0,0,0}\makebox(0,0)[b]{\smash{3.7}}}%
    \put(0.21218412,0.04784434){\color[rgb]{0,0,0}\makebox(0,0)[b]{\smash{3.8}}}%
    \put(0.31233985,0.04784434){\color[rgb]{0,0,0}\makebox(0,0)[b]{\smash{3.9}}}%
    \put(0.4123945,0.04784434){\color[rgb]{0,0,0}\makebox(0,0)[b]{\smash{4.0}}}%
    \put(0.51257897,0.04784434){\color[rgb]{0,0,0}\makebox(0,0)[b]{\smash{4.1}}}%
    \put(0.3373301,0.00370933){\color[rgb]{0,0,0}\makebox(0,0)[b]{\smash{Switching current, $I_{sw}$ ($\upmu$A)}}}%
    \put(0.06884316,0.34066681){\color[rgb]{0,0,0}\makebox(0,0)[b]{\smash{400}}}%
    \put(0.06884316,0.27537597){\color[rgb]{0,0,0}\makebox(0,0)[b]{\smash{300}}}%
    \put(0.06884316,0.21008928){\color[rgb]{0,0,0}\makebox(0,0)[b]{\smash{200}}}%
    \put(0.07972871,0.0755276){\color[rgb]{0,0,0}\makebox(0,0)[b]{\smash{0}}}%
    \put(0.06884316,0.14460558){\color[rgb]{0,0,0}\makebox(0,0)[b]{\smash{100}}}%
    \put(0.06884316,0.40617112){\color[rgb]{0,0,0}\makebox(0,0)[b]{\smash{500}}}%
    \put(0.00163065,0.45958407){\color[rgb]{0,0,0}\makebox(0,0)[lb]{\smash{\textbf{a)}}}}%
    \put(0.57059649,0.45958407){\color[rgb]{0,0,0}\makebox(0,0)[lb]{\smash{\textbf{b)}}}}%
    \put(0.01583028,0.26345914){\color[rgb]{0,0,0}\rotatebox{90}{\makebox(0,0)[b]{\smash{Counts}}}}%
    \put(0.66762149,0.12182982){\color[rgb]{1,1,1}\makebox(0,0)[b]{\smash{200 nm}}}%
    \put(0.94261181,0.17588913){\color[rgb]{0,0,0}\makebox(0,0)[lb]{\smash{90 nm}}}%
  \end{picture}%
\endgroup%

%% file: All.bbl
\begin{thebibliography}{37}%
\makeatletter
\providecommand \@ifxundefined [1]{%
 \@ifx{#1\undefined}
}%
\providecommand \@ifnum [1]{%
 \ifnum #1\expandafter \@firstoftwo
 \else \expandafter \@secondoftwo
 \fi
}%
\providecommand \@ifx [1]{%
 \ifx #1\expandafter \@firstoftwo
 \else \expandafter \@secondoftwo
 \fi
}%
\providecommand \natexlab [1]{#1}%
\providecommand \enquote  [1]{``#1''}%
\providecommand \bibnamefont  [1]{#1}%
\providecommand \bibfnamefont [1]{#1}%
\providecommand \citenamefont [1]{#1}%
\providecommand \href@noop [0]{\@secondoftwo}%
\providecommand \href [0]{\begingroup \@sanitize@url \@href}%
\providecommand \@href[1]{\@@startlink{#1}\@@href}%
\providecommand \@@href[1]{\endgroup#1\@@endlink}%
\providecommand \@sanitize@url [0]{\catcode `\\12\catcode `\$12\catcode
  `\&12\catcode `\#12\catcode `\^12\catcode `\_12\catcode `\%12\relax}%
\providecommand \@@startlink[1]{}%
\providecommand \@@endlink[0]{}%
\providecommand \url  [0]{\begingroup\@sanitize@url \@url }%
\providecommand \@url [1]{\endgroup\@href {#1}{\urlprefix }}%
\providecommand \urlprefix  [0]{URL }%
\providecommand \Eprint [0]{\href }%
\providecommand \doibase [0]{http://dx.doi.org/}%
\providecommand \selectlanguage [0]{\@gobble}%
\providecommand \bibinfo  [0]{\@secondoftwo}%
\providecommand \bibfield  [0]{\@secondoftwo}%
\providecommand \translation [1]{[#1]}%
\providecommand \BibitemOpen [0]{}%
\providecommand \bibitemStop [0]{}%
\providecommand \bibitemNoStop [0]{.\EOS\space}%
\providecommand \EOS [0]{\spacefactor3000\relax}%
\providecommand \BibitemShut  [1]{\csname bibitem#1\endcsname}%
\let\auto@bib@innerbib\@empty
\bibitem [{\citenamefont {Anderson}(1959)}]{ANDERSON1959}%
  \BibitemOpen
  \bibfield  {author} {\bibinfo {author} {\bibfnamefont {P.}~\bibnamefont
  {Anderson}},\ }\href {\doibase https://doi.org/10.1016/0022-3697(59)90036-8}
  {\bibfield  {journal} {\bibinfo  {journal} {J. Phys. Chem. Solids}\ }\textbf
  {\bibinfo {volume} {11}},\ \bibinfo {pages} {26 } (\bibinfo {year}
  {1959})}\BibitemShut {NoStop}%
\bibitem [{\citenamefont {Beloborodov}\ \emph {et~al.}(2007)\citenamefont
  {Beloborodov}, \citenamefont {Lopatin}, \citenamefont {Vinokur},\ and\
  \citenamefont {Efetov}}]{Efetov2007}%
  \BibitemOpen
  \bibfield  {author} {\bibinfo {author} {\bibfnamefont {I.~S.}\ \bibnamefont
  {Beloborodov}}, \bibinfo {author} {\bibfnamefont {A.~V.}\ \bibnamefont
  {Lopatin}}, \bibinfo {author} {\bibfnamefont {V.~M.}\ \bibnamefont
  {Vinokur}}, \ and\ \bibinfo {author} {\bibfnamefont {K.~B.}\ \bibnamefont
  {Efetov}},\ }\href {\doibase 10.1103/RevModPhys.79.469} {\bibfield  {journal}
  {\bibinfo  {journal} {Rev. Mod. Phys.}\ }\textbf {\bibinfo {volume} {79}},\
  \bibinfo {pages} {469} (\bibinfo {year} {2007})}\BibitemShut {NoStop}%
\bibitem [{\citenamefont {Cohen}\ and\ \citenamefont
  {Abeles}(1968)}]{Cohen1968}%
  \BibitemOpen
  \bibfield  {author} {\bibinfo {author} {\bibfnamefont {R.~W.}\ \bibnamefont
  {Cohen}}\ and\ \bibinfo {author} {\bibfnamefont {B.}~\bibnamefont {Abeles}},\
  }\href {\doibase 10.1103/PhysRev.168.444} {\bibfield  {journal} {\bibinfo
  {journal} {Phys. Rev.}\ }\textbf {\bibinfo {volume} {168}},\ \bibinfo {pages}
  {444} (\bibinfo {year} {1968})}\BibitemShut {NoStop}%
\bibitem [{\citenamefont {Deutscher}\ \emph
  {et~al.}(1973{\natexlab{a}})\citenamefont {Deutscher}, \citenamefont
  {Fenichel}, \citenamefont {Gershenson}, \citenamefont {Gr{\"u}nbaum},\ and\
  \citenamefont {Ovadyahu}}]{Deutscher1973}%
  \BibitemOpen
  \bibfield  {author} {\bibinfo {author} {\bibfnamefont {G.}~\bibnamefont
  {Deutscher}}, \bibinfo {author} {\bibfnamefont {H.}~\bibnamefont {Fenichel}},
  \bibinfo {author} {\bibfnamefont {M.}~\bibnamefont {Gershenson}}, \bibinfo
  {author} {\bibfnamefont {E.}~\bibnamefont {Gr{\"u}nbaum}}, \ and\ \bibinfo
  {author} {\bibfnamefont {Z.}~\bibnamefont {Ovadyahu}},\ }\href {\doibase
  10.1007/BF00655256} {\bibfield  {journal} {\bibinfo  {journal} {J. Low Temp.
  Phys.}\ }\textbf {\bibinfo {volume} {10}},\ \bibinfo {pages} {231} (\bibinfo
  {year} {1973}{\natexlab{a}})}\BibitemShut {NoStop}%
\bibitem [{\citenamefont {Parmenter}(1967)}]{Parmenter1967}%
  \BibitemOpen
  \bibfield  {author} {\bibinfo {author} {\bibfnamefont {R.~H.}\ \bibnamefont
  {Parmenter}},\ }\href {\doibase 10.1103/PhysRev.154.353} {\bibfield
  {journal} {\bibinfo  {journal} {Phys. Rev.}\ }\textbf {\bibinfo {volume}
  {154}},\ \bibinfo {pages} {353} (\bibinfo {year} {1967})}\BibitemShut
  {NoStop}%
\bibitem [{\citenamefont {Dynes}\ and\ \citenamefont
  {Garno}(1981)}]{Dynes1981}%
  \BibitemOpen
  \bibfield  {author} {\bibinfo {author} {\bibfnamefont {R.~C.}\ \bibnamefont
  {Dynes}}\ and\ \bibinfo {author} {\bibfnamefont {J.~P.}\ \bibnamefont
  {Garno}},\ }\href {\doibase 10.1103/PhysRevLett.46.137} {\bibfield  {journal}
  {\bibinfo  {journal} {Phys. Rev. Lett.}\ }\textbf {\bibinfo {volume} {46}},\
  \bibinfo {pages} {137} (\bibinfo {year} {1981})}\BibitemShut {NoStop}%
\bibitem [{\citenamefont {Pracht}\ \emph {et~al.}(2016)\citenamefont {Pracht},
  \citenamefont {Bachar}, \citenamefont {Benfatto}, \citenamefont {Deutscher},
  \citenamefont {Farber}, \citenamefont {Dressel},\ and\ \citenamefont
  {Scheffler}}]{Pracht2016}%
  \BibitemOpen
  \bibfield  {author} {\bibinfo {author} {\bibfnamefont {U.~S.}\ \bibnamefont
  {Pracht}}, \bibinfo {author} {\bibfnamefont {N.}~\bibnamefont {Bachar}},
  \bibinfo {author} {\bibfnamefont {L.}~\bibnamefont {Benfatto}}, \bibinfo
  {author} {\bibfnamefont {G.}~\bibnamefont {Deutscher}}, \bibinfo {author}
  {\bibfnamefont {E.}~\bibnamefont {Farber}}, \bibinfo {author} {\bibfnamefont
  {M.}~\bibnamefont {Dressel}}, \ and\ \bibinfo {author} {\bibfnamefont
  {M.}~\bibnamefont {Scheffler}},\ }\href {\doibase 10.1103/PhysRevB.93.100503}
  {\bibfield  {journal} {\bibinfo  {journal} {Phys. Rev. B}\ }\textbf {\bibinfo
  {volume} {93}},\ \bibinfo {pages} {100503} (\bibinfo {year}
  {2016})}\BibitemShut {NoStop}%
\bibitem [{\citenamefont {Abeles}\ \emph {et~al.}(1966)\citenamefont {Abeles},
  \citenamefont {Cohen},\ and\ \citenamefont {Cullen}}]{Abeles1966}%
  \BibitemOpen
  \bibfield  {author} {\bibinfo {author} {\bibfnamefont {B.}~\bibnamefont
  {Abeles}}, \bibinfo {author} {\bibfnamefont {R.~W.}\ \bibnamefont {Cohen}}, \
  and\ \bibinfo {author} {\bibfnamefont {G.~W.}\ \bibnamefont {Cullen}},\
  }\href {\doibase 10.1103/PhysRevLett.17.632} {\bibfield  {journal} {\bibinfo
  {journal} {Phys. Rev. Lett.}\ }\textbf {\bibinfo {volume} {17}},\ \bibinfo
  {pages} {632} (\bibinfo {year} {1966})}\BibitemShut {NoStop}%
\bibitem [{\citenamefont {Deutscher}\ and\ \citenamefont
  {Dodds}(1977)}]{Deutscher1977}%
  \BibitemOpen
  \bibfield  {author} {\bibinfo {author} {\bibfnamefont {G.}~\bibnamefont
  {Deutscher}}\ and\ \bibinfo {author} {\bibfnamefont {S.~A.}\ \bibnamefont
  {Dodds}},\ }\href {\doibase 10.1103/PhysRevB.16.3936} {\bibfield  {journal}
  {\bibinfo  {journal} {Phys. Rev. B}\ }\textbf {\bibinfo {volume} {16}},\
  \bibinfo {pages} {3936} (\bibinfo {year} {1977})}\BibitemShut {NoStop}%
\bibitem [{\citenamefont {Chui}\ \emph {et~al.}(1981)\citenamefont {Chui},
  \citenamefont {Lindenfeld}, \citenamefont {McLean},\ and\ \citenamefont
  {Mui}}]{Chui1981}%
  \BibitemOpen
  \bibfield  {author} {\bibinfo {author} {\bibfnamefont {T.}~\bibnamefont
  {Chui}}, \bibinfo {author} {\bibfnamefont {P.}~\bibnamefont {Lindenfeld}},
  \bibinfo {author} {\bibfnamefont {W.~L.}\ \bibnamefont {McLean}}, \ and\
  \bibinfo {author} {\bibfnamefont {K.}~\bibnamefont {Mui}},\ }\href {\doibase
  10.1103/PhysRevB.24.6728} {\bibfield  {journal} {\bibinfo  {journal} {Phys.
  Rev. B}\ }\textbf {\bibinfo {volume} {24}},\ \bibinfo {pages} {6728}
  (\bibinfo {year} {1981})}\BibitemShut {NoStop}%
\bibitem [{\citenamefont {Rotzinger}\ \emph {et~al.}(2016)\citenamefont
  {Rotzinger}, \citenamefont {Skacel}, \citenamefont {Pfirrmann}, \citenamefont
  {Voss}, \citenamefont {Münzberg}, \citenamefont {Probst}, \citenamefont
  {Bushev}, \citenamefont {Weides}, \citenamefont {Ustinov},\ and\
  \citenamefont {Mooij}}]{Rotzinger2016}%
  \BibitemOpen
  \bibfield  {author} {\bibinfo {author} {\bibfnamefont {H.}~\bibnamefont
  {Rotzinger}}, \bibinfo {author} {\bibfnamefont {S.~T.}\ \bibnamefont
  {Skacel}}, \bibinfo {author} {\bibfnamefont {M.}~\bibnamefont {Pfirrmann}},
  \bibinfo {author} {\bibfnamefont {J.~N.}\ \bibnamefont {Voss}}, \bibinfo
  {author} {\bibfnamefont {J.}~\bibnamefont {Münzberg}}, \bibinfo {author}
  {\bibfnamefont {S.}~\bibnamefont {Probst}}, \bibinfo {author} {\bibfnamefont
  {P.}~\bibnamefont {Bushev}}, \bibinfo {author} {\bibfnamefont {M.~P.}\
  \bibnamefont {Weides}}, \bibinfo {author} {\bibfnamefont {A.~V.}\
  \bibnamefont {Ustinov}}, \ and\ \bibinfo {author} {\bibfnamefont {J.~E.}\
  \bibnamefont {Mooij}},\ }\href@noop {} {\bibfield  {journal} {\bibinfo
  {journal} {Supercond. Sci. Technol.}\ }\textbf {\bibinfo {volume} {30}}
  (\bibinfo {year} {2016})}\BibitemShut {NoStop}%
\bibitem [{\citenamefont {Wallraff}\ \emph {et~al.}(2004)\citenamefont
  {Wallraff}, \citenamefont {Schuster}, \citenamefont {Blais}, \citenamefont
  {Frunzio}, \citenamefont {Huang}, \citenamefont {Majer}, \citenamefont
  {Kumar}, \citenamefont {Girvin},\ and\ \citenamefont
  {Schoelkopf}}]{Wallraff2004}%
  \BibitemOpen
  \bibfield  {author} {\bibinfo {author} {\bibfnamefont {A.}~\bibnamefont
  {Wallraff}}, \bibinfo {author} {\bibfnamefont {D.~I.}\ \bibnamefont
  {Schuster}}, \bibinfo {author} {\bibfnamefont {A.}~\bibnamefont {Blais}},
  \bibinfo {author} {\bibfnamefont {L.}~\bibnamefont {Frunzio}}, \bibinfo
  {author} {\bibfnamefont {R.-S.}\ \bibnamefont {Huang}}, \bibinfo {author}
  {\bibfnamefont {J.}~\bibnamefont {Majer}}, \bibinfo {author} {\bibfnamefont
  {S.}~\bibnamefont {Kumar}}, \bibinfo {author} {\bibfnamefont {S.~M.}\
  \bibnamefont {Girvin}}, \ and\ \bibinfo {author} {\bibfnamefont {R.~J.}\
  \bibnamefont {Schoelkopf}},\ }\href {http://dx.doi.org/10.1038/nature02851}
  {\bibfield  {journal} {\bibinfo  {journal} {Nature}\ }\textbf {\bibinfo
  {volume} {431}},\ \bibinfo {pages} {162} (\bibinfo {year}
  {2004})}\BibitemShut {NoStop}%
\bibitem [{\citenamefont {Gu}\ \emph {et~al.}(2017)\citenamefont {Gu},
  \citenamefont {Kockum}, \citenamefont {Miranowicz}, \citenamefont {xi~Liu},\
  and\ \citenamefont {Nori}}]{Nori2017}%
  \BibitemOpen
  \bibfield  {author} {\bibinfo {author} {\bibfnamefont {X.}~\bibnamefont
  {Gu}}, \bibinfo {author} {\bibfnamefont {A.~F.}\ \bibnamefont {Kockum}},
  \bibinfo {author} {\bibfnamefont {A.}~\bibnamefont {Miranowicz}}, \bibinfo
  {author} {\bibfnamefont {Y.}~\bibnamefont {xi~Liu}}, \ and\ \bibinfo {author}
  {\bibfnamefont {F.}~\bibnamefont {Nori}},\ }\href@noop {} {\bibfield
  {journal} {\bibinfo  {journal} {Physics Reports}\ }\textbf {\bibinfo {volume}
  {718–719}},\ \bibinfo {pages} {1} (\bibinfo {year} {2017})}\BibitemShut
  {NoStop}%
\bibitem [{\citenamefont {Day}\ \emph {et~al.}(2003)\citenamefont {Day},
  \citenamefont {LeDuc}, \citenamefont {Mazin}, \citenamefont {Vayonakis},\
  and\ \citenamefont {Zmuidzinas}}]{Day2003}%
  \BibitemOpen
  \bibfield  {author} {\bibinfo {author} {\bibfnamefont {P.~K.}\ \bibnamefont
  {Day}}, \bibinfo {author} {\bibfnamefont {H.~G.}\ \bibnamefont {LeDuc}},
  \bibinfo {author} {\bibfnamefont {B.~A.}\ \bibnamefont {Mazin}}, \bibinfo
  {author} {\bibfnamefont {A.}~\bibnamefont {Vayonakis}}, \ and\ \bibinfo
  {author} {\bibfnamefont {J.}~\bibnamefont {Zmuidzinas}},\ }\href
  {http://dx.doi.org/10.1038/nature02037} {\bibfield  {journal} {\bibinfo
  {journal} {Nature}\ }\textbf {\bibinfo {volume} {425}},\ \bibinfo {pages}
  {817} (\bibinfo {year} {2003})}\BibitemShut {NoStop}%
\bibitem [{\citenamefont {Emery}\ and\ \citenamefont
  {Kivelson}(1995)}]{Emery1995}%
  \BibitemOpen
  \bibfield  {author} {\bibinfo {author} {\bibfnamefont {V.~J.}\ \bibnamefont
  {Emery}}\ and\ \bibinfo {author} {\bibfnamefont {S.~A.}\ \bibnamefont
  {Kivelson}},\ }\href {http://dx.doi.org/10.1038/374434a0} {\bibfield
  {journal} {\bibinfo  {journal} {Nature}\ }\textbf {\bibinfo {volume} {374}},\
  \bibinfo {pages} {434} (\bibinfo {year} {1995})}\BibitemShut {NoStop}%
\bibitem [{\citenamefont {Deutscher}\ \emph
  {et~al.}(1973{\natexlab{b}})\citenamefont {Deutscher}, \citenamefont
  {Gershenson}, \citenamefont {Grünbaum},\ and\ \citenamefont
  {Imry}}]{Deutscher1973-2}%
  \BibitemOpen
  \bibfield  {author} {\bibinfo {author} {\bibfnamefont {G.}~\bibnamefont
  {Deutscher}}, \bibinfo {author} {\bibfnamefont {M.}~\bibnamefont
  {Gershenson}}, \bibinfo {author} {\bibfnamefont {E.}~\bibnamefont
  {Grünbaum}}, \ and\ \bibinfo {author} {\bibfnamefont {Y.}~\bibnamefont
  {Imry}},\ }\href {\doibase 10.1116/1.1318416} {\bibfield  {journal} {\bibinfo
   {journal} {Journal of Vacuum Science and Technology}\ }\textbf {\bibinfo
  {volume} {10}},\ \bibinfo {pages} {697} (\bibinfo {year}
  {1973}{\natexlab{b}})},\ \Eprint
  {http://arxiv.org/abs/https://doi.org/10.1116/1.1318416}
  {https://doi.org/10.1116/1.1318416} \BibitemShut {NoStop}%
\bibitem [{\citenamefont {Pracht}\ \emph {et~al.}(2017)\citenamefont {Pracht},
  \citenamefont {Cea}, \citenamefont {Bachar}, \citenamefont {Deutscher},
  \citenamefont {Farber}, \citenamefont {Dressel}, \citenamefont {Scheffler},
  \citenamefont {Castellani}, \citenamefont {Garc\'{\i}a-Garc\'{\i}a},\ and\
  \citenamefont {Benfatto}}]{Pracht2017}%
  \BibitemOpen
  \bibfield  {author} {\bibinfo {author} {\bibfnamefont {U.~S.}\ \bibnamefont
  {Pracht}}, \bibinfo {author} {\bibfnamefont {T.}~\bibnamefont {Cea}},
  \bibinfo {author} {\bibfnamefont {N.}~\bibnamefont {Bachar}}, \bibinfo
  {author} {\bibfnamefont {G.}~\bibnamefont {Deutscher}}, \bibinfo {author}
  {\bibfnamefont {E.}~\bibnamefont {Farber}}, \bibinfo {author} {\bibfnamefont
  {M.}~\bibnamefont {Dressel}}, \bibinfo {author} {\bibfnamefont
  {M.}~\bibnamefont {Scheffler}}, \bibinfo {author} {\bibfnamefont
  {C.}~\bibnamefont {Castellani}}, \bibinfo {author} {\bibfnamefont {A.~M.}\
  \bibnamefont {Garc\'{\i}a-Garc\'{\i}a}}, \ and\ \bibinfo {author}
  {\bibfnamefont {L.}~\bibnamefont {Benfatto}},\ }\href {\doibase
  10.1103/PhysRevB.96.094514} {\bibfield  {journal} {\bibinfo  {journal} {Phys.
  Rev. B}\ }\textbf {\bibinfo {volume} {96}},\ \bibinfo {pages} {094514}
  (\bibinfo {year} {2017})}\BibitemShut {NoStop}%
\bibitem [{\citenamefont {Bachar}\ \emph
  {et~al.}(2015{\natexlab{a}})\citenamefont {Bachar}, \citenamefont {Lerer},
  \citenamefont {Levy}, \citenamefont {Hacohen-Gourgy}, \citenamefont {Almog},
  \citenamefont {Saadaoui}, \citenamefont {Salman}, \citenamefont {Morenzoni},\
  and\ \citenamefont {Deutscher}}]{Bachar2015}%
  \BibitemOpen
  \bibfield  {author} {\bibinfo {author} {\bibfnamefont {N.}~\bibnamefont
  {Bachar}}, \bibinfo {author} {\bibfnamefont {S.}~\bibnamefont {Lerer}},
  \bibinfo {author} {\bibfnamefont {A.}~\bibnamefont {Levy}}, \bibinfo {author}
  {\bibfnamefont {S.}~\bibnamefont {Hacohen-Gourgy}}, \bibinfo {author}
  {\bibfnamefont {B.}~\bibnamefont {Almog}}, \bibinfo {author} {\bibfnamefont
  {H.}~\bibnamefont {Saadaoui}}, \bibinfo {author} {\bibfnamefont
  {Z.}~\bibnamefont {Salman}}, \bibinfo {author} {\bibfnamefont
  {E.}~\bibnamefont {Morenzoni}}, \ and\ \bibinfo {author} {\bibfnamefont
  {G.}~\bibnamefont {Deutscher}},\ }\href {\doibase 10.1103/PhysRevB.91.041123}
  {\bibfield  {journal} {\bibinfo  {journal} {Phys. Rev. B}\ }\textbf {\bibinfo
  {volume} {91}},\ \bibinfo {pages} {041123} (\bibinfo {year}
  {2015}{\natexlab{a}})}\BibitemShut {NoStop}%
\bibitem [{\citenamefont {Bachar}\ \emph
  {et~al.}(2015{\natexlab{b}})\citenamefont {Bachar}, \citenamefont {Pracht},
  \citenamefont {Farber}, \citenamefont {Dressel}, \citenamefont {Deutscher},\
  and\ \citenamefont {Scheffler}}]{BacharPracht2015}%
  \BibitemOpen
  \bibfield  {author} {\bibinfo {author} {\bibfnamefont {N.}~\bibnamefont
  {Bachar}}, \bibinfo {author} {\bibfnamefont {U.~S.}\ \bibnamefont {Pracht}},
  \bibinfo {author} {\bibfnamefont {E.}~\bibnamefont {Farber}}, \bibinfo
  {author} {\bibfnamefont {M.}~\bibnamefont {Dressel}}, \bibinfo {author}
  {\bibfnamefont {G.}~\bibnamefont {Deutscher}}, \ and\ \bibinfo {author}
  {\bibfnamefont {M.}~\bibnamefont {Scheffler}},\ }\href {\doibase
  10.1007/s10909-014-1244-z} {\bibfield  {journal} {\bibinfo  {journal} {J. Low
  Temp. Phys.}\ }\textbf {\bibinfo {volume} {179}},\ \bibinfo {pages} {83}
  (\bibinfo {year} {2015}{\natexlab{b}})}\BibitemShut {NoStop}%
\bibitem [{\citenamefont {Astafiev}\ \emph {et~al.}(2012)\citenamefont
  {Astafiev}, \citenamefont {Ioffe}, \citenamefont {Kafanov}, \citenamefont
  {Pashkin}, \citenamefont {Arutyunov}, \citenamefont {Shahar}, \citenamefont
  {Cohen},\ and\ \citenamefont {Tsai}}]{Astafiev2012}%
  \BibitemOpen
  \bibfield  {author} {\bibinfo {author} {\bibfnamefont {O.~V.}\ \bibnamefont
  {Astafiev}}, \bibinfo {author} {\bibfnamefont {L.~B.}\ \bibnamefont {Ioffe}},
  \bibinfo {author} {\bibfnamefont {S.}~\bibnamefont {Kafanov}}, \bibinfo
  {author} {\bibfnamefont {Y.~A.}\ \bibnamefont {Pashkin}}, \bibinfo {author}
  {\bibfnamefont {K.~Y.}\ \bibnamefont {Arutyunov}}, \bibinfo {author}
  {\bibfnamefont {D.}~\bibnamefont {Shahar}}, \bibinfo {author} {\bibfnamefont
  {O.}~\bibnamefont {Cohen}}, \ and\ \bibinfo {author} {\bibfnamefont {J.~S.}\
  \bibnamefont {Tsai}},\ }\href {http://dx.doi.org/10.1038/nature10930}
  {\bibfield  {journal} {\bibinfo  {journal} {Nature}\ }\textbf {\bibinfo
  {volume} {484}},\ \bibinfo {pages} {355} (\bibinfo {year}
  {2012})}\BibitemShut {NoStop}%
\bibitem [{\citenamefont {Manucharyan}\ \emph {et~al.}(2009)\citenamefont
  {Manucharyan}, \citenamefont {Koch}, \citenamefont {Glazman},\ and\
  \citenamefont {Devoret}}]{Manucharyan2009}%
  \BibitemOpen
  \bibfield  {author} {\bibinfo {author} {\bibfnamefont {V.~E.}\ \bibnamefont
  {Manucharyan}}, \bibinfo {author} {\bibfnamefont {J.}~\bibnamefont {Koch}},
  \bibinfo {author} {\bibfnamefont {L.~I.}\ \bibnamefont {Glazman}}, \ and\
  \bibinfo {author} {\bibfnamefont {M.~H.}\ \bibnamefont {Devoret}},\ }\href
  {\doibase 10.1126/science.1175552} {\bibfield  {journal} {\bibinfo  {journal}
  {Science}\ }\textbf {\bibinfo {volume} {326}},\ \bibinfo {pages} {113}
  (\bibinfo {year} {2009})}\BibitemShut {NoStop}%
\bibitem [{\citenamefont {Gladchenko}\ \emph {et~al.}(2008)\citenamefont
  {Gladchenko}, \citenamefont {Olaya}, \citenamefont {Dupont-Ferrier},
  \citenamefont {Douçot}, \citenamefont {Ioffe},\ and\ \citenamefont
  {Gershenson}}]{Gladchenko2008}%
  \BibitemOpen
  \bibfield  {author} {\bibinfo {author} {\bibfnamefont {S.}~\bibnamefont
  {Gladchenko}}, \bibinfo {author} {\bibfnamefont {D.}~\bibnamefont {Olaya}},
  \bibinfo {author} {\bibfnamefont {E.}~\bibnamefont {Dupont-Ferrier}},
  \bibinfo {author} {\bibfnamefont {B.}~\bibnamefont {Douçot}}, \bibinfo
  {author} {\bibfnamefont {L.~B.}\ \bibnamefont {Ioffe}}, \ and\ \bibinfo
  {author} {\bibfnamefont {M.~E.}\ \bibnamefont {Gershenson}},\ }\href
  {http://dx.doi.org/10.1038/nphys1151} {\bibfield  {journal} {\bibinfo
  {journal} {Nature Physics}\ }\textbf {\bibinfo {volume} {5}},\ \bibinfo
  {pages} {48} (\bibinfo {year} {2008})}\BibitemShut {NoStop}%
\bibitem [{\citenamefont {Ho~Eom}\ \emph {et~al.}(2012)\citenamefont {Ho~Eom},
  \citenamefont {Day}, \citenamefont {LeDuc},\ and\ \citenamefont
  {Zmuidzinas}}]{Eom2012}%
  \BibitemOpen
  \bibfield  {author} {\bibinfo {author} {\bibfnamefont {B.}~\bibnamefont
  {Ho~Eom}}, \bibinfo {author} {\bibfnamefont {P.~K.}\ \bibnamefont {Day}},
  \bibinfo {author} {\bibfnamefont {H.~G.}\ \bibnamefont {LeDuc}}, \ and\
  \bibinfo {author} {\bibfnamefont {J.}~\bibnamefont {Zmuidzinas}},\ }\href
  {http://dx.doi.org/10.1038/nphys2356} {\bibfield  {journal} {\bibinfo
  {journal} {Nat. Phys.}\ }\textbf {\bibinfo {volume} {8}},\ \bibinfo {pages}
  {623} (\bibinfo {year} {2012})}\BibitemShut {NoStop}%
\bibitem [{\citenamefont {Puri}\ \emph {et~al.}(2017)\citenamefont {Puri},
  \citenamefont {Boutin},\ and\ \citenamefont {Blais}}]{Puri2017}%
  \BibitemOpen
  \bibfield  {author} {\bibinfo {author} {\bibfnamefont {S.}~\bibnamefont
  {Puri}}, \bibinfo {author} {\bibfnamefont {S.}~\bibnamefont {Boutin}}, \ and\
  \bibinfo {author} {\bibfnamefont {A.}~\bibnamefont {Blais}},\ }\href
  {https://doi.org/10.1038/s41534-017-0019-1} {\bibfield  {journal} {\bibinfo
  {journal} {Quantum Information}\ }\textbf {\bibinfo {volume} {3}},\ \bibinfo
  {pages} {18} (\bibinfo {year} {2017})}\BibitemShut {NoStop}%
\bibitem [{\citenamefont {D.F.~Walls}(2008)}]{QO2008}%
  \BibitemOpen
  \bibfield  {author} {\bibinfo {author} {\bibfnamefont {G.~J.~M.}\
  \bibnamefont {D.F.~Walls}},\ }\href@noop {} {\emph {\bibinfo {title} {Quantum
  Optics}}}\ (\bibinfo  {publisher} {Springer-Verlag Berlin Heidelberg},\
  \bibinfo {year} {2008})\BibitemShut {NoStop}%
\bibitem [{\citenamefont {Martin}\ and\ \citenamefont
  {Puplett}(1970)}]{MARTIN1970}%
  \BibitemOpen
  \bibfield  {author} {\bibinfo {author} {\bibfnamefont {D.}~\bibnamefont
  {Martin}}\ and\ \bibinfo {author} {\bibfnamefont {E.}~\bibnamefont
  {Puplett}},\ }\href {\doibase http://dx.doi.org/10.1016/0020-0891(70)90006-0}
  {\bibfield  {journal} {\bibinfo  {journal} {Infrared Phys.}\ }\textbf
  {\bibinfo {volume} {10}},\ \bibinfo {pages} {105 } (\bibinfo {year}
  {1970})}\BibitemShut {NoStop}%
\bibitem [{\citenamefont {{Catalano, A.}}\ \emph {et~al.}(2015)\citenamefont
  {{Catalano, A.}}, \citenamefont {{Goupy, J.}}, \citenamefont {{le Sueur,
  H.}}, \citenamefont {{Benoit, A.}}, \citenamefont {{Bourrion, O.}},
  \citenamefont {{Calvo, M.}}, \citenamefont {{D’addabbo, A.}}, \citenamefont
  {{Dumoulin, L.}}, \citenamefont {{Levy-Bertrand, F.}}, \citenamefont
  {{Macías-Pérez, J.}}, \citenamefont {{Marnieros, S.}}, \citenamefont
  {{Ponthieu, N.}},\ and\ \citenamefont {{Monfardini, A.}}}]{Calvo2015}%
  \BibitemOpen
  \bibfield  {author} {\bibinfo {author} {\bibnamefont {{Catalano, A.}}},
  \bibinfo {author} {\bibnamefont {{Goupy, J.}}}, \bibinfo {author}
  {\bibnamefont {{le Sueur, H.}}}, \bibinfo {author} {\bibnamefont {{Benoit,
  A.}}}, \bibinfo {author} {\bibnamefont {{Bourrion, O.}}}, \bibinfo {author}
  {\bibnamefont {{Calvo, M.}}}, \bibinfo {author} {\bibnamefont {{D’addabbo,
  A.}}}, \bibinfo {author} {\bibnamefont {{Dumoulin, L.}}}, \bibinfo {author}
  {\bibnamefont {{Levy-Bertrand, F.}}}, \bibinfo {author} {\bibnamefont
  {{Macías-Pérez, J.}}}, \bibinfo {author} {\bibnamefont {{Marnieros, S.}}},
  \bibinfo {author} {\bibnamefont {{Ponthieu, N.}}}, \ and\ \bibinfo {author}
  {\bibnamefont {{Monfardini, A.}}},\ }\href {\doibase
  10.1051/0004-6361/201526206} {\bibfield  {journal} {\bibinfo  {journal}
  {A\&A}\ }\textbf {\bibinfo {volume} {580}},\ \bibinfo {pages} {A15} (\bibinfo
  {year} {2015})}\BibitemShut {NoStop}%
\bibitem [{\citenamefont {Hutter}\ \emph {et~al.}(2011)\citenamefont {Hutter},
  \citenamefont {Thol\'en}, \citenamefont {Stannigel}, \citenamefont {Lidmar},\
  and\ \citenamefont {Haviland}}]{Hutter2011}%
  \BibitemOpen
  \bibfield  {author} {\bibinfo {author} {\bibfnamefont {C.}~\bibnamefont
  {Hutter}}, \bibinfo {author} {\bibfnamefont {E.~A.}\ \bibnamefont
  {Thol\'en}}, \bibinfo {author} {\bibfnamefont {K.}~\bibnamefont {Stannigel}},
  \bibinfo {author} {\bibfnamefont {J.}~\bibnamefont {Lidmar}}, \ and\ \bibinfo
  {author} {\bibfnamefont {D.~B.}\ \bibnamefont {Haviland}},\ }\href {\doibase
  10.1103/PhysRevB.83.014511} {\bibfield  {journal} {\bibinfo  {journal} {Phys.
  Rev. B}\ }\textbf {\bibinfo {volume} {83}},\ \bibinfo {pages} {014511}
  (\bibinfo {year} {2011})}\BibitemShut {NoStop}%
\bibitem [{\citenamefont {Wei\ss{}l}\ \emph {et~al.}(2015)\citenamefont
  {Wei\ss{}l}, \citenamefont {K\"ung}, \citenamefont {Dumur}, \citenamefont
  {Feofanov}, \citenamefont {Matei}, \citenamefont {Naud}, \citenamefont
  {Buisson}, \citenamefont {Hekking},\ and\ \citenamefont
  {Guichard}}]{Weissl2015}%
  \BibitemOpen
  \bibfield  {author} {\bibinfo {author} {\bibfnamefont {T.}~\bibnamefont
  {Wei\ss{}l}}, \bibinfo {author} {\bibfnamefont {B.}~\bibnamefont {K\"ung}},
  \bibinfo {author} {\bibfnamefont {E.}~\bibnamefont {Dumur}}, \bibinfo
  {author} {\bibfnamefont {A.~K.}\ \bibnamefont {Feofanov}}, \bibinfo {author}
  {\bibfnamefont {I.}~\bibnamefont {Matei}}, \bibinfo {author} {\bibfnamefont
  {C.}~\bibnamefont {Naud}}, \bibinfo {author} {\bibfnamefont {O.}~\bibnamefont
  {Buisson}}, \bibinfo {author} {\bibfnamefont {F.~W.~J.}\ \bibnamefont
  {Hekking}}, \ and\ \bibinfo {author} {\bibfnamefont {W.}~\bibnamefont
  {Guichard}},\ }\href {\doibase 10.1103/PhysRevB.92.104508} {\bibfield
  {journal} {\bibinfo  {journal} {Phys. Rev. B}\ }\textbf {\bibinfo {volume}
  {92}},\ \bibinfo {pages} {104508} (\bibinfo {year} {2015})}\BibitemShut
  {NoStop}%
\bibitem [{\citenamefont {Bourassa}\ \emph {et~al.}(2012)\citenamefont
  {Bourassa}, \citenamefont {Beaudoin}, \citenamefont {Gambetta},\ and\
  \citenamefont {Blais}}]{Bourassa2012}%
  \BibitemOpen
  \bibfield  {author} {\bibinfo {author} {\bibfnamefont {J.}~\bibnamefont
  {Bourassa}}, \bibinfo {author} {\bibfnamefont {F.}~\bibnamefont {Beaudoin}},
  \bibinfo {author} {\bibfnamefont {J.~M.}\ \bibnamefont {Gambetta}}, \ and\
  \bibinfo {author} {\bibfnamefont {A.}~\bibnamefont {Blais}},\ }\href
  {\doibase 10.1103/PhysRevA.86.013814} {\bibfield  {journal} {\bibinfo
  {journal} {Phys. Rev. A}\ }\textbf {\bibinfo {volume} {86}},\ \bibinfo
  {pages} {013814} (\bibinfo {year} {2012})}\BibitemShut {NoStop}%
\bibitem [{\citenamefont {Masluk}\ \emph {et~al.}(2012)\citenamefont {Masluk},
  \citenamefont {Pop}, \citenamefont {Kamal}, \citenamefont {Minev},\ and\
  \citenamefont {Devoret}}]{Masluk2012}%
  \BibitemOpen
  \bibfield  {author} {\bibinfo {author} {\bibfnamefont {N.~A.}\ \bibnamefont
  {Masluk}}, \bibinfo {author} {\bibfnamefont {I.~M.}\ \bibnamefont {Pop}},
  \bibinfo {author} {\bibfnamefont {A.}~\bibnamefont {Kamal}}, \bibinfo
  {author} {\bibfnamefont {Z.~K.}\ \bibnamefont {Minev}}, \ and\ \bibinfo
  {author} {\bibfnamefont {M.~H.}\ \bibnamefont {Devoret}},\ }\href {\doibase
  10.1103/PhysRevLett.109.137002} {\bibfield  {journal} {\bibinfo  {journal}
  {Phys. Rev. Lett.}\ }\textbf {\bibinfo {volume} {109}},\ \bibinfo {pages}
  {137002} (\bibinfo {year} {2012})}\BibitemShut {NoStop}%
\bibitem [{\citenamefont {Grunhaupt}\ \emph {et~al.}(2018)\citenamefont
  {Grunhaupt}, \citenamefont {Maleeva}, \citenamefont {Skacel}, \citenamefont
  {Levy-Bertrand}, \citenamefont {Ustinov}, \citenamefont {Rotzinger},
  \citenamefont {Monfardini},\ and\ \citenamefont {Pop}}]{Lukas}%
  \BibitemOpen
  \bibfield  {author} {\bibinfo {author} {\bibfnamefont {L.}~\bibnamefont
  {Grunhaupt}}, \bibinfo {author} {\bibfnamefont {N.}~\bibnamefont {Maleeva}},
  \bibinfo {author} {\bibfnamefont {S.~T.}\ \bibnamefont {Skacel}}, \bibinfo
  {author} {\bibfnamefont {F.}~\bibnamefont {Levy-Bertrand}}, \bibinfo {author}
  {\bibfnamefont {A.~V.}\ \bibnamefont {Ustinov}}, \bibinfo {author}
  {\bibfnamefont {H.}~\bibnamefont {Rotzinger}}, \bibinfo {author}
  {\bibfnamefont {A.}~\bibnamefont {Monfardini}}, \ and\ \bibinfo {author}
  {\bibfnamefont {I.~M.}\ \bibnamefont {Pop}},\ }\href@noop {} {\bibfield
  {journal} {\bibinfo  {journal} {arXiv:1802.01858}\ } (\bibinfo {year}
  {2018})}\BibitemShut {NoStop}%
\bibitem [{\citenamefont {Cardani}\ \emph {et~al.}(2017)\citenamefont
  {Cardani}, \citenamefont {Bellini}, \citenamefont {Casali}, \citenamefont
  {Castellano}, \citenamefont {Colantoni}, \citenamefont {Coppolecchia},
  \citenamefont {Cosmelli}, \citenamefont {Cruciani}, \citenamefont
  {D'Addabbo}, \citenamefont {Domizio}, \citenamefont {Martinez}, \citenamefont
  {Tomei},\ and\ \citenamefont {Vignati}}]{CARDANI2017}%
  \BibitemOpen
  \bibfield  {author} {\bibinfo {author} {\bibfnamefont {L.}~\bibnamefont
  {Cardani}}, \bibinfo {author} {\bibfnamefont {F.}~\bibnamefont {Bellini}},
  \bibinfo {author} {\bibfnamefont {N.}~\bibnamefont {Casali}}, \bibinfo
  {author} {\bibfnamefont {M.}~\bibnamefont {Castellano}}, \bibinfo {author}
  {\bibfnamefont {I.}~\bibnamefont {Colantoni}}, \bibinfo {author}
  {\bibfnamefont {A.}~\bibnamefont {Coppolecchia}}, \bibinfo {author}
  {\bibfnamefont {C.}~\bibnamefont {Cosmelli}}, \bibinfo {author}
  {\bibfnamefont {A.}~\bibnamefont {Cruciani}}, \bibinfo {author}
  {\bibfnamefont {A.}~\bibnamefont {D'Addabbo}}, \bibinfo {author}
  {\bibfnamefont {S.~D.}\ \bibnamefont {Domizio}}, \bibinfo {author}
  {\bibfnamefont {M.}~\bibnamefont {Martinez}}, \bibinfo {author}
  {\bibfnamefont {C.}~\bibnamefont {Tomei}}, \ and\ \bibinfo {author}
  {\bibfnamefont {M.}~\bibnamefont {Vignati}},\ }\href {\doibase
  https://doi.org/10.1016/j.nima.2016.04.011} {\bibfield  {journal} {\bibinfo
  {journal} {Nucl. Instrum. Meth. Phys. Res., Sect. A}\ }\textbf {\bibinfo
  {volume} {845}},\ \bibinfo {pages} {338 } (\bibinfo {year} {2017})},\
  \bibinfo {note} {proceedings of the Vienna Conference on Instrumentation
  2016}\BibitemShut {NoStop}%
\bibitem [{\citenamefont {Devoret}\ \emph {et~al.}(2004)\citenamefont
  {Devoret}, \citenamefont {Wallraff},\ and\ \citenamefont
  {Martinis}}]{Devoret2004}%
  \BibitemOpen
  \bibfield  {author} {\bibinfo {author} {\bibfnamefont {M.~H.}\ \bibnamefont
  {Devoret}}, \bibinfo {author} {\bibfnamefont {A.}~\bibnamefont {Wallraff}}, \
  and\ \bibinfo {author} {\bibfnamefont {J.~M.}\ \bibnamefont {Martinis}},\
  }\href@noop {} {\bibfield  {journal} {\bibinfo  {journal}
  {arXiv:cond-mat/0411174}\ } (\bibinfo {year} {2004})}\BibitemShut {NoStop}%
\bibitem [{\citenamefont {Probst}\ \emph {et~al.}(2015)\citenamefont {Probst},
  \citenamefont {Song}, \citenamefont {Bushev}, \citenamefont {Ustinov},\ and\
  \citenamefont {Weides}}]{Probst2015}%
  \BibitemOpen
  \bibfield  {author} {\bibinfo {author} {\bibfnamefont {S.}~\bibnamefont
  {Probst}}, \bibinfo {author} {\bibfnamefont {F.~B.}\ \bibnamefont {Song}},
  \bibinfo {author} {\bibfnamefont {P.~A.}\ \bibnamefont {Bushev}}, \bibinfo
  {author} {\bibfnamefont {A.~V.}\ \bibnamefont {Ustinov}}, \ and\ \bibinfo
  {author} {\bibfnamefont {M.}~\bibnamefont {Weides}},\ }\href {\doibase
  10.1063/1.4907935} {\bibfield  {journal} {\bibinfo  {journal} {Rev. Sci.
  Instrum.}\ }\textbf {\bibinfo {volume} {86}},\ \bibinfo {pages} {024706}
  (\bibinfo {year} {2015})},\ \Eprint
  {http://arxiv.org/abs/http://dx.doi.org/10.1063/1.4907935}
  {http://dx.doi.org/10.1063/1.4907935} \BibitemShut {NoStop}%
\bibitem [{\citenamefont {Bluhm}(2008)}]{bluhm2008deconvolution}%
  \BibitemOpen
  \bibfield  {author} {\bibinfo {author} {\bibfnamefont {D.}~\bibnamefont
  {Bluhm}},\ }\emph {\bibinfo {title} {A deconvolution method for switching
  current histograms as a fast diagnosis tool}},\ \href@noop {} {Ph.D.
  thesis},\ \bibinfo  {school} {The Graduate School, Stony Brook University:
  Stony Brook, NY.} (\bibinfo {year} {2008})\BibitemShut {NoStop}%
\bibitem [{\citenamefont {Kurkij{\"a}rvi}(1972)}]{kurkijarvi1972intrinsic}%
  \BibitemOpen
  \bibfield  {author} {\bibinfo {author} {\bibfnamefont {J.}~\bibnamefont
  {Kurkij{\"a}rvi}},\ }\href@noop {} {\bibfield  {journal} {\bibinfo  {journal}
  {Phys. Rev. B}\ }\textbf {\bibinfo {volume} {6}},\ \bibinfo {pages} {832}
  (\bibinfo {year} {1972})}\BibitemShut {NoStop}%
\end{thebibliography}%
